\pdfoutput=1
\documentclass[11pt]{article}

\usepackage[preprint]{acl}

\usepackage{times}
\usepackage{latexsym}

\usepackage[T1]{fontenc}

\usepackage[utf8]{inputenc}

\usepackage{microtype}

\usepackage{inconsolata}
\usepackage{dashundergaps}
\usepackage{graphicx}
\usepackage{algorithm}
\usepackage{algorithmic}
\usepackage{multirow}
\usepackage{booktabs}
\usepackage{colortbl}
\usepackage{amsmath}
\usepackage{amssymb}
\usepackage{amsfonts}
\usepackage{makecell}
\usepackage{newfloat}
\usepackage{listings}
\usepackage{lipsum}
\usepackage{nicefrac}
\usepackage{array}
\usepackage{enumitem}
\usepackage{float}
\usepackage{placeins}
\usepackage{threeparttable}
\setlist[itemize]{nosep, leftmargin=*}
\usepackage{tabularx}
\usepackage[most]{tcolorbox}
\tcbuselibrary{listings,breakable,skins}
\usepackage{listings}
\usepackage{xcolor}
\definecolor{thirdblue}{RGB}{41, 128, 185} 
\newcommand{\third}[1]{\textcolor{thirdblue}{#1}}
\lstdefinestyle{promptstyle}{
    basicstyle=\ttfamily\scriptsize,
    columns=fullflexible,
    keepspaces=true,
    breaklines=true,
    breakatwhitespace=false,
    breakautoindent=false,
    breakindent=0pt,
    postbreak={},
    showstringspaces=false,
    upquote=true,
    tabsize=2,
    numbers=none,
    xleftmargin=0pt,
    framexleftmargin=0pt
}

\newtcblisting{promptbox}[2][]{
    enhanced,
    breakable,
    listing only,
    listing engine=listings,
    colback=gray!3,
    colframe=gray!55,
    coltitle=black,
    fonttitle=\bfseries,
    title={#2},
    listing options={
        style=promptstyle,
        gobble=4
    },
    before skip=6pt,
    after skip=6pt,
    left=1mm,
    right=1mm,
    top=1mm,
    bottom=1mm,
    #1
}
\newcolumntype{C}{>{\centering\arraybackslash}X} 

\title{Spatial-Omni: Spatial Audio Understanding Integration \\ in Multimodal LLMs via FOA Encoding}

\author{
Zhiyuan Zhu\textsuperscript{1,2}\quad
Yixuan Chen\textsuperscript{1}\quad
Yiwen Shao\textsuperscript{2}\quad
Wenxiang Guo\textsuperscript{1} \\
\textbf{Changhao Pan\textsuperscript{1}} \quad
\textbf{Yu Zhang \quad}
\textbf{Yuxiang Wang\textsuperscript{2}\quad}
\textbf{Wei Liu\textsuperscript{2}\quad} \\
\textbf{Houhua Zhang\textsuperscript{1}\quad}
\textbf{Chengkuan Zeng\textsuperscript{1} \quad}
\textbf{Wenbo Cheng\textsuperscript{1}\quad}
\textbf{Yunxi Liu\textsuperscript{1}\quad}\\
\textbf{Rui Yang\textsuperscript{1}\quad}
\textbf{Steve Yves\textsuperscript{2}\quad}
\textbf{Liefeng Bo\textsuperscript{2}\quad}
\textbf{Zhou Zhao\textsuperscript{1, $\dagger$}}\\
\vspace{0.1cm}
\textsuperscript{1}Zhejiang University\quad
\textsuperscript{2}Tencent Hunyuan\\
\texttt{schmittzhu@zju.edu.cn}
}

\begin{document}
\maketitle

\begingroup
\def\thefootnote{}
\footnotetext{Work done during internship at Tencent Hunyuan with project leader Yiwen Shao. 
$\dagger$\ Corresponding author to Zhou Zhao <zhaozhou@zju.edu.cn> }
\endgroup

\begin{abstract}
Recent multimodal large language models mainly process audio as monaural signals, thereby discarding the spatial cues contained in spatial audio for sound localization, spatial relation reasoning, and spatial scene understanding.
We propose \textbf{Spatial-Omni}, a lightweight method that implements \textbf{SO-Encoder} to inject First-Order Ambisonics (FOA) spatial audio into existing Multimodal LLMs or Large-Audio Language Models (LALMs) as an independent modality, without modifying their original audio encoders.
SO-Encoder provides spatial tokens with limited additional context cost and improves spatial audio understanding through efficient staged training.
To support training and evaluation, we construct \textbf{SO-Dataset}, \textbf{SO-QA}, and \textbf{SO-Bench} from open-source data, real recordings, and simulations, containing 400K FOA spatial audio clips and 2.1M spatial question-answering pairs.
SO-Bench covers 16 spatial audio understanding subtasks, including basic detection and location estimation, spatial relation understanding, and complex spatial reasoning.
Experiments show that Spatial-Omni outperforms existing open-source LALMs and Omni LLMs on most spatial audio understanding subtasks; general audio understanding degrades without the original training data, but lightweight mixed training partially recovers it.
Code and data are available at \url{https://github.com/dieKarotte/Spatial-Omni}.
\end{abstract}

\section{Introduction}
\label{sec:intro}

Spatial audio preserves directional, distance, motion, and environmental cues beyond sound content, providing the basis for 3D auditory scene understanding \cite{zhu2025asaudio, lu2025deep}.
These cues have been studied in tasks such as sound event localization and detection, source motion analysis, and spatial scene reasoning \cite{adavanne2018sound, aparicio2024baseline}.
However, most current LALMs and Omni LLMs process audio as monaural signals, so their audio inputs collapse spatial cues before language-model reasoning.

Representative audio and Omni models \cite{ghosh2024gama, tang2023salmonn, ghosh2026audio, xu2025qwen25omnitechnicalreport, chu2024qwen2, xu2025qwen3, team2026qwen3, ding2025kimi} commonly rely on pretrained audio encoders, such as Whisper \cite{radford2023robust} or Audio Spectrogram Transformers \cite{gong2021ast}, to extract acoustic features for language-model reasoning.
This design provides strong semantic audio understanding, but the audio pathway is usually optimized for monaural content and does not explicitly preserve spatial structure.
Recent spatial audio LLMs \cite{zeng2026listen, mishra2025spatial, dementyev2026phasecoder, jiang2026sci, sridhar2026spatial, sakshi2025spur, tang2024can} show that spatial cues can benefit LLM-based audio reasoning, but many of them retrain or heavily modify the original audio encoder to accept spatial input.
Retraining risks the semantic ability that the base encoder was pretrained for, and a retrained encoder no longer serves monaural and spatial audio equally well.
A second obstacle is data: large-scale FOA spatial audio QA corpora and systematic benchmarks remain scarce, so spatial reasoning ability is hard to train and evaluate in a controlled way.

To address these limitations, we propose \textbf{Spatial-Omni}, a framework for injecting spatial audio understanding into Omni LLMs through an independent spatial modality.
Its core component is \textbf{SO-Encoder}, a lightweight spatial encoder added in parallel to the original audio encoder.
Given FOA input $(W, Y, Z, X)$, the original Omni audio encoder continues to receive the W channel and preserves the base model's audio semantic pathway.
In parallel, SO-Encoder extracts spatial cues from FOA mel features and Intensity Vector (IV) features, including direction, distance, motion, and multi-source spatial relations.
A Temporal Pixel Shuffle Projector then compresses the frame-level spatial latents and maps them into compact spatial tokens, allowing the LLM to jointly attend to audio, spatial, visual, and text tokens.
This design upgrades spatial understanding without replacing the original audio pathway, and stays compatible with the abilities of existing Omni LLMs.

We further construct the data and evaluation pipeline needed for this setting.
\textbf{SO-Dataset} contains about 400K FOA spatial audio clips collected from public sound event detection and localization (SELD) datasets, real-world recordings, and simulation.
Based on their metadata, we build \textbf{SO-QA}, a large-scale spatial audio question-answering dataset with 2.1M QA pairs covering source detection, time localization, direction and distance estimation, spatial relation understanding, motion analysis, multi-source reasoning, and spatial speech recognition.
For evaluation, we introduce \textbf{SO-Bench}, a benchmark with 16 spatial audio subtasks grouped into basic detection and estimation, spatial relation understanding, and complex reasoning with semantics.
We train SO-Encoder on SO-Dataset using SELD supervision, and further fine-tune Spatial-Omni upgraded Omni LLMs using SO-QA.
Experiments compare general LALMs, Omni LLMs, spatial audio baselines, and Spatial-Omni variants.
On SO-Bench, Spatial-Omni improves most spatial tasks by clear margins, and the ablations attribute the gains to real spatial tokens rather than the token interface alone.
Our contributions are as follows:
\begin{itemize}[leftmargin=*]
\item Spatial-Omni and SO-Encoder: a lightweight spatial encoding branch that injects FOA spatial audio into LALMs and Omni LLMs as an independent modality, with the base audio encoder left untouched.
\item SO-Dataset \& SO-QA: about 400K FOA audio clips and 2.1M spatial QA pairs built from public data, real recordings, and simulations.
\item SO-Bench: a 16-task benchmark that spans basic localization, spatial relation reasoning, and complex semantic tasks, on which we evaluate a broad set of existing LALMs and Omni LLMs.
\end{itemize}

\begin{figure*}[htbp]
    \centering
    \includegraphics[width=\textwidth]{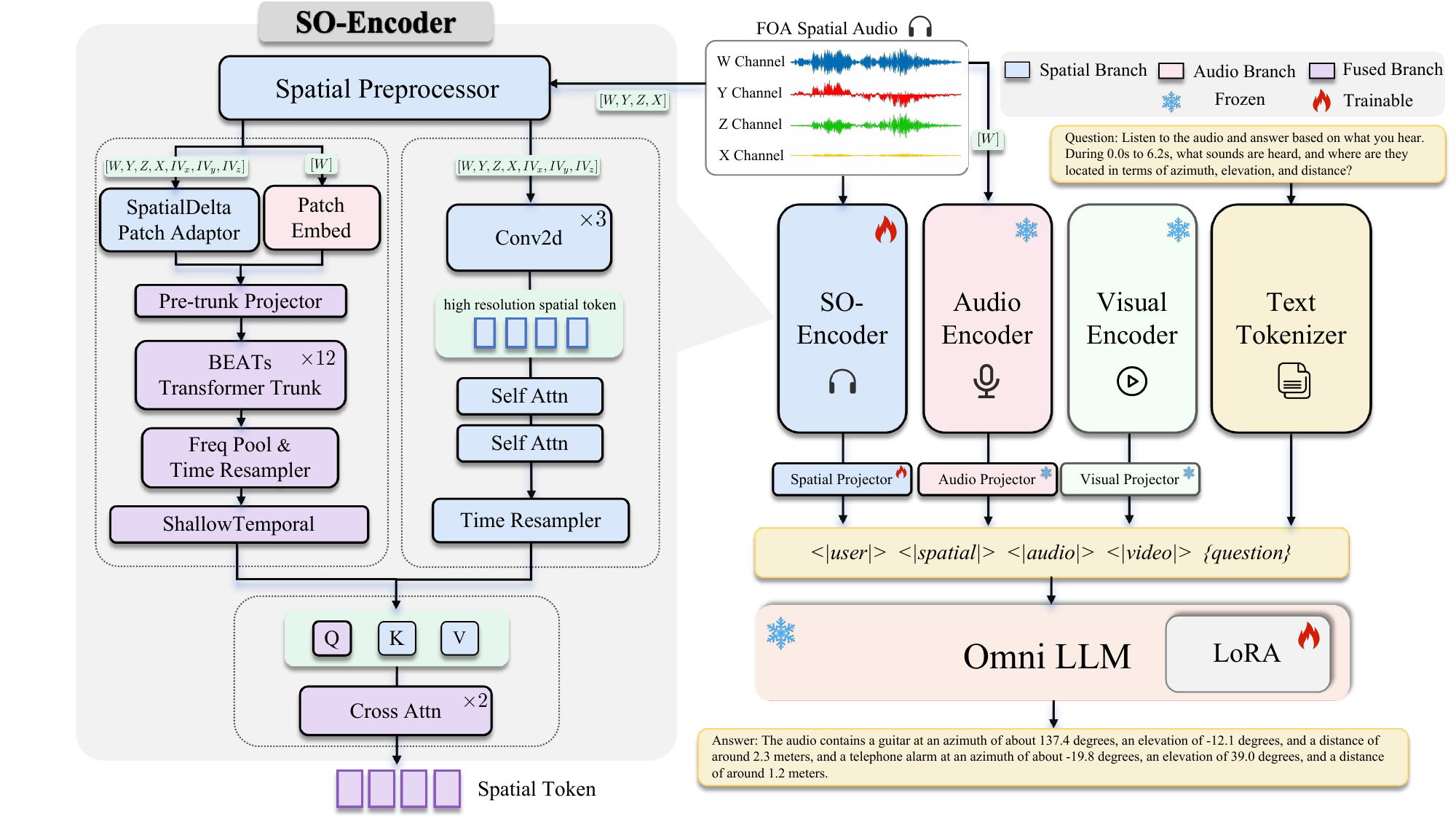}
    \caption{The overall architecture of the proposed Spatial-Omni. Details of SO-Encoder are shown in the left box. The original audio encoder is kept unchanged to preserve the base model's semantic ability, while the parallel SO-Encoder extracts spatial cues from FOA features. A lightweight projector maps the spatial latents into the LLM token space for joint learning with audio, visual, and text tokens.}
    \label{fig:model}
\end{figure*}

\section{Related Work}
\label{sec:rel}

\subsection{Spatial Audio}
Spatial audio represents 3D auditory scenes by preserving direction, distance, motion, and environmental cues beyond sound content.
Common formats include binaural audio, multichannel audio, and Ambisonics.
Recent spatial audio research can be viewed from two connected directions: understanding and generation \cite{zhu2025asaudio, lu2025deep}.
For understanding, earlier SELD tasks \cite{aparicio2024baseline, hu2025pseldnets} focus on event classification and direction estimation.
Recent work expands this scope to motion tracking, source separation, audio-visual scene analysis, and spatial acoustic consistency modeling.
It also studies spatial representation learning and language alignment, which connect low-level spatial cues with semantic decoding and text-based retrieval \cite{hu2025salm, wilkinghoff2026dspast, paik2026natural, seki2026spatial, sudarsanam2025towards}.
For generation, recent methods have moved from upmixing and visually guided spatialization to neural generative models \cite{DBLP:conf/icce-tw/YoshidaTOH23, morgadoNIPS18, sun2024both, zhang2025isdrama, leng2022binauralgrad, kushwaha2025diff, liu2025omniaudiogeneratingspatialaudio}.
Spatial cues are becoming a central variable in audio and multimodal modeling.
This motivates LALMs and Omni LLMs to reason and understand spatial audio rather than only recognize sound events.

\subsection{LALMs, Omni LLMs and Spatial Audio LLMs}
LALMs connect audio encoders with language models for natural-language interaction over acoustic content, covering audio-text alignment, audio question answering, and acoustic reasoning \cite{elizalde2023clap, tang2023salmonn, ghosh2024gama, deshmukh2023pengi, chu2024qwen2, kong2024audio, ding2025kimi}.
Recent Omni models further integrate audio, vision, and language, providing a stronger foundation for general multimodal understanding \cite{ghosh2026audio,  abouelenin2025phi4mini, xu2025qwen25omnitechnicalreport, comanici2025gemini, achiam2023gpt}.
However, these models do not explicitly use spatial audio to model direction, distance, motion, or inter-source spatial relations needed for spatial reasoning.

Existing Spatial Audio LLMs extend spatial audio understanding into the LLM framework by adding spatial encoders or spatial features to LLMs.
Binaural audio methods build explicit spatial encoders \cite{zheng2024bat}, add visual context \cite{anonymous2026hear, zeng2026listen}, condition on Direction of Arrival (DoA), or model room impulse responses (RIRs) \cite{mishra2025spatial, biswas2025owl} to support spatial question answering and acoustic scene reasoning, showing that explicit spatial cues help LLMs understand acoustic scenes.
Studies on FOA or multichannel input inject IV features into the base audio encoder or modify the original encoder to extract stronger spectral-spatial representations \cite{tang2024can, sakshi2025spur}.
Others train a parallel spatial encoder \cite{jiang2026sci}, learn neural IV features \cite{liu2026jaeger}, adapt to different microphone geometries \cite{dementyev2026phasecoder}, or fuse semantic and spatial experts to reduce mono bias \cite{you2026world}.
These studies broaden Spatial Audio LLMs from binaural perception to sound-field input and format generalization.
However, binding spatial modeling to substantial modification or retraining of the original audio encoder may affect the semantic ability of the original encoder and complicate joint support for monaural audio and spatial audio in a single audio encoder.

A gap remains in the evaluation of spatial audio understanding.
General Multimodal LLM benchmarks and visual spatial benchmarks focus on spatial cognition from visual inputs rather than spatial audio \cite{yu2023mm, yu2024mm, yue2024mmmu, liu2024mmbench, azuma2022scanqa}.
Recent audio benchmarks begin to test binaural motion, audio-visual viewpoint reasoning, and spatial tasks \cite{sun2025spatial, sridhar2026spatial, chen2026savvy, liu2025star, kumar2026mmau}.
However, they still provide limited coverage of FOA audio, multi-source spatial relations, motion analysis, and comprehensive spatial question answering.
Additional details on Spatial Audio LLMs and benchmarks are provided in Appendix~\ref{app:relatedwork_details}.
\section{Method}
\label{sec:method}

\paragraph{Model Overview}
\label{sec:model}
Figure~\ref{fig:model} shows the overall architecture.
We use Qwen Omni series model as the base Omni LLM and add a parallel SO-Encoder beside the original audio and visual encoders.
We further conduct more experiments on LALMs to verify the generalization of our method and build audio omni capabilities.
For FOA input $(W, Y, Z, X)$, the W channel is sent to the original audio encoder to preserve the semantic ability of the base model.
SO-Encoder receives 4 FOA mel features and 3 Intensity Vector (IV) features and outputs frame-level spatial latents at 10 Hz.
A lightweight projector compresses and maps these latents into the LLM token space.
The LLM then jointly attends to audio, spatial, visual, and text tokens.
This design treats spatial audio as an independent modality while keeping the model flexible for both single-channel audio and spatial audio without modifying the original audio encoder.

\paragraph{SO-Encoder}
\label{sec:spatial_encoder}
It extracts a residual spatial delta from the seven-channel input with three-stage convolution, maps it to the patch embedding dimension, and adds it to the pretrained BEATs W channel patch embedding with a learnable weight $\alpha$.
This injects spatial information while minimizing damage to the original BEATs semantic representation.
The latent then passes through ShallowTemporal module, using a one-layer Transformer with LayerNorm to produce the semantic tokens for fusion.

The spatial branch uses a CNN and Transformer structure with three 2D CNN layers and two Transformer layers.
It extracts high-resolution features for DoA, distance, motion, and overlapping sources before resampled to lower frame level spatial tokens.
LocalSpatialCrossFuser then fuses the two branches with two layers of gated cross attention.
Semantic tokens are queries, and spatial tokens are keys and values.
A sigmoid gate controls the fusion ratio and outputs the final spatial token.

For supervision, we add an event head to support event F-score.
We use a two-stage SourceQueryDecoder with $K$ track queries.
The heads predict activity, class, direction vector, and distance with softplus.
During inference, Hungarian matching assigns predictions to stable tracks across frames.

\paragraph{Projector}
\label{sec:projector}
After the SO-Encoder outputs 10 Hz features, a lightweight projector maps them to the LLM space and compresses time.
Temporal Pixel Shuffle Projector downsamples the one-dimensional token sequence.

Given encoder output $\mathbf{Z} \in \mathbb{R}^{B \times T \times d_s}$, we divide time into non-overlapping groups of $k$ frames and concatenate them along the feature dimension:
$
\hat{\mathbf{Z}} = \mathrm{Shuffle}_k(\mathbf{Z}) \in \mathbb{R}^{B \times \lfloor T/k \rfloor \times k d_s}
$
This reduces the temporal length from $T$ to $\lfloor T/k \rfloor$ while preserving local frame information through feature concatenation.
We then use a two-layer MLP with LayerNorm to project $\hat{\mathbf{Z}}$ into the LLM hidden dimension $d_{\ell}$, producing $\mathbf{S} \in \mathbb{R}^{B \times \lfloor T/k \rfloor \times d_\ell}$.
This provides compact spatial tokens with limited LLM context cost.

\section{Dataset and Benchmark}
\label{sec:dataset}

\begin{figure*}[t]
    \centering
    \includegraphics[width=\textwidth]{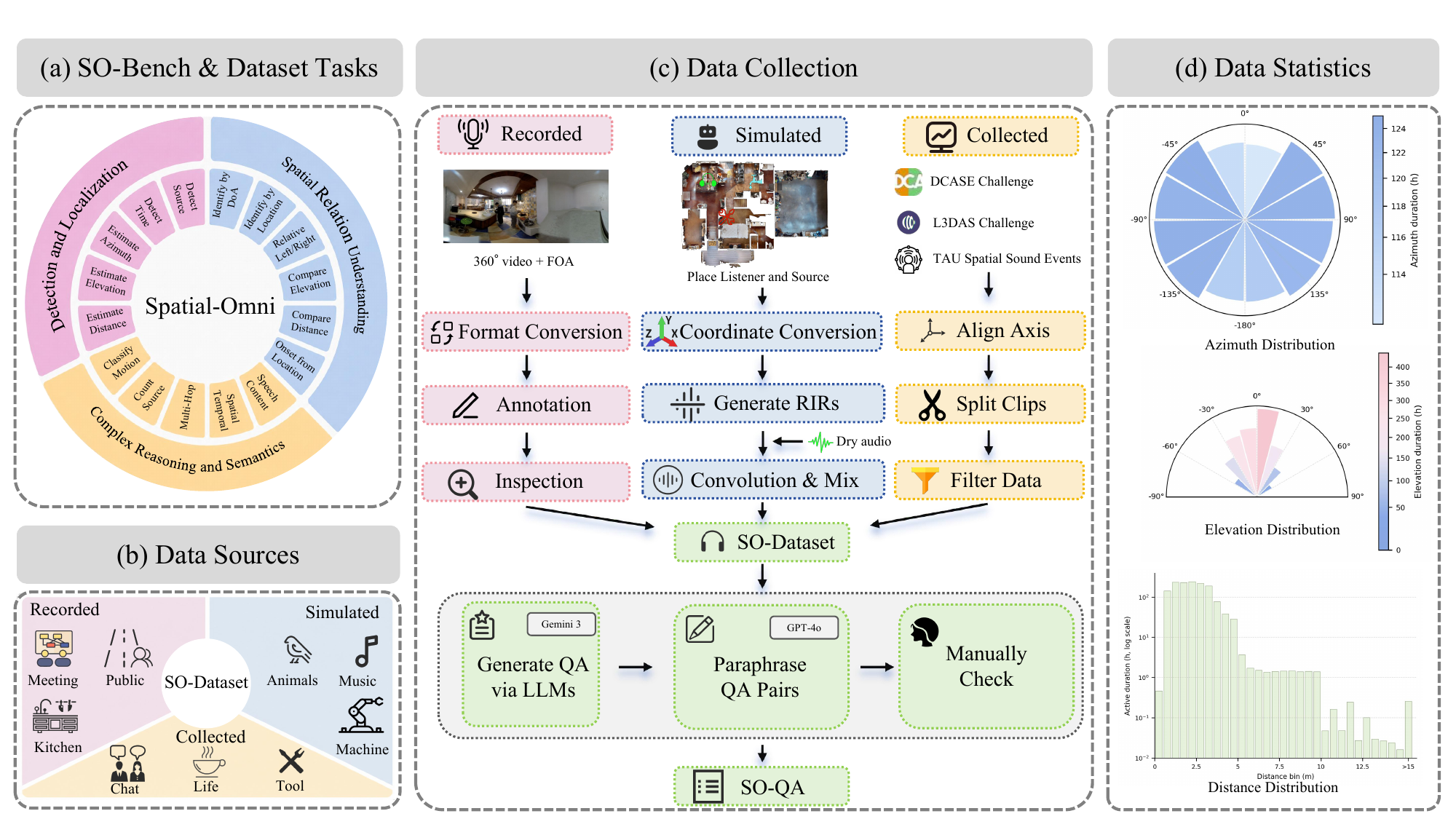}
    \caption{(a) The sub-tasks of our SO-Bench, (b) The data sources in our SO-Dataset, (c) The data collection pipeline of SO-Dataset \& SO-QA, and (d) The statistical distribution of our SO-Dataset.}
    \label{fig:data}
\end{figure*}

There are only limited open-source QA datasets for spatial audio.
We construct SO-Dataset and SO-QA for spatial audio training, and SO-Bench for evaluation.
The data covers open-source annotations, real recordings, and simulation, which improves generalization across scenes, sources, and dynamic spatial reasoning.
SO-Dataset, SO-QA and SO-Bench statistics are shown in Figure~\ref{fig:data}.

\paragraph{SO-Dataset and SO-QA}
\label{sec:spatial_qa_dataset}
SO-Dataset contains FOA spatial audio from open-source datasets, real recordings, and simulations, covering indoor and outdoor scenes, diverse spatial distributions, and overlapping sound sources.
The overall dataset contains 63 event classes and about 400K FOA clips, where each clip contains one or more overlapping sound events.
The open-source data includes public SELD datasets L3DAS22, 23 \cite{guizzo2022l3das22, marinoni2023overview}, TAU Spatial Sound Events 2019, 2020, 2021 \cite{adavanne2019tau, adavanne2019moving, politis2020taunigens, politis2021taunigens}, and STARSS22, STARSS23 \cite{politis2022starss22,shimada2023starss23}.
We unify their annotation formats and class labels, resulting in 27.8k clips.
The recorded data covers indoor events and outdoor events, with 3.5k clips from 23 scenes paired with 360-degree visual data.
The simulated data is generated with SoundSpace 2.0 \cite{chen2022soundspaces} using HM3D, MP3D \cite{ramakrishnan2021habitat}, and Replica \cite{straub2019replica} rooms with semantic annotations.
We randomly sample listener position, listener orientation, and source position, and generate static \& dynamic RIRs.
Dry audio events from FSD50K \cite{fonseca2021fsd50k} and speech from LibriSpeech \cite{panayotov2015librispeech} are convolved with the simulated RIRs, and then mixed to create overlapping sound event audio clips.
The simulated part contains about 370k FOA clips.
The test rooms do not overlap with training rooms, enabling evaluation on unseen rooms.
Metadata of SO-Dataset includes event class, active interval, frame-level azimuth, elevation, distance for each sound source, as well as motion information for dynamic sources.    

Based on metadata of SO-Dataset, we construct SO-QA for Spatial Audio LLM training and evaluation.
For each sub-task, we first manually design 20-25 question templates and answer templates, then instantiate QA pairs with metadata and audio using Gemini-3, and use GPT-4o to paraphrase them for language diversity.
SO-QA contains about 2.1M spatial question answering pairs.

\paragraph{SO-Bench Design}
\label{sec:benchmark}
We design SO-Bench to evaluate basic detection and estimation, spatial relation understanding, and complex reasoning with semantics.
It uses an independent spatial QA test set for fair evaluation and contains 7k QA pairs.

Basic detection and estimation tasks include Detect Source (DS), Detect Time (DT), Estimation of Azimuth (EAzi), Estimation of Elevation (EEle), and Estimation of Distance (EDis).
They evaluate source detection, event timing, and the estimation of source direction and distance.
Spatial relation understanding tasks include Identify Source by DoA (IS-DoA), Identify Source by Location (IS-Loc), Relative Left-Right (RLR), Comparison between Elevation (CEle), Comparison between Distance (CDis), and Onset from Location (OL).
They evaluate whether a model can identify the correct source from direction or location descriptions and compare spatial relations between sources.
Although OL outputs temporal information, it first requires the model to associate a given spatial location with the correct source.
Complex reasoning and semantics tasks include Classify Motion (CM), Count Sources (CS), Multi Hop (MH), Spatial Temporal Caption (ST), and Speech Content (SC).
They test dynamic event understanding, source counting, multi-hop spatial reasoning, global spatial-temporal scene understanding, and speech recognition under spatial conditions.
Additional details are provided in Appendix~\ref{appendix:benchmark}.

\section{Experiment}
\subsection{Training Details}
\label{sec:training_details}

\paragraph{SO-Encoder}
\label{sec:spatial_encoder_training}
The SO-Encoder is trained with SELD losses: a 63-way cross-entropy for event class, a Top-$K$ ranking loss with a weak BCE anchor for track activity, a cosine loss on unit direction vectors for DoA, and a smooth-$\ell_1$ loss for distance, with predicted and ground-truth tracks aligned by Hungarian matching.
Semantic class learning and spatial feature learning compete early in training: when both losses are backpropagated together, the spatial branch can disturb the event classification ability of the BEATs semantic branch.
Following the training experience of BAT, we adopt a two-stage strategy that first stabilizes class learning and then introduces spatial learning.

\paragraph{Spatial-Omni Models}
\label{sec:spatial_llm_training}
Spatial LLM training follows a three-stage strategy.
In the first stage, we connect the SO-Encoder but freeze the base LLM and the main body of SO-Encoder.
Only the projector and necessary adapters are trained, so that projected spatial tokens align with the LLM hidden space.
In the second stage, we enable LLM LoRA \cite{hu2022lora} and jointly train it with the projector.
This teaches the LLM to use spatial tokens to answer spatial questions in SO-QA.
In the third stage, we unfreeze the trainable parts of the SO-Encoder.
The SO-Encoder, projector, and LLM LoRA are then jointly adapted with the QA loss.
This gradual unfreezing strategy reduces the interference of spatial features with the original model's ability early in training.

\paragraph{Training and Hyperparameters}
\label{sec:hyperparameters}
All training uses the AdamW optimizer with cosine decay learning rate scheduling, and FOA audio is sampled at 16 kHz.
Both SO-Encoder and models with 4B/7B base LLM are trained on 8$\times$NVIDIA A100 GPUs, while SO-30B, the version based on Qwen3-Omni-30B-A3B Instruct \cite{xu2025qwen3}, is trained on 8$\times$H20 GPUs.
More details on training and hyperparameters are provided in Appendix~\ref{app: training_details}.

\subsection{Evaluation Metrics}
\label{sec:evaluation_metrics}
For SO-Encoder, we report F20 (correct sound event classification with DoA error below 20 degrees) for correctness of both sound event detection and localization, DoA error for azimuth and elevation localization, and relative distance error for source distance estimation.

For SO-Bench, metrics are chosen according to task type.
DS uses event F1, and DT uses median time-span IoU.
EAzi, EEle, and EDis use accuracy within an angular or distance tolerance, together with angle error or absolute distance error for auxiliary analysis.
IS-DoA and IS-Loc evaluate whether the model can identify the correct source from direction or location descriptions.
Classification tasks, including RLR, CEle, CDis, CM, CS, and MH, use accuracy in percentage.
Comparison and open-ended answers such as OL and ST are first parsed with rules; an LLM judge then resolves the rest, accepting a comparison only when the compared entities and the direction both match the reference.
We use Word Error Rate (WER) to evaluate spatial speech content recognition in the SC task.
All models use greedy decoding in the main experiments so that the comparison remains fair across all evaluated open-source baselines.

\subsection{Baselines}
\label{sec:baselines}
We use the open-source DCASE 2024 baseline as the main encoder baseline, and also include the best and second-place challenge systems, whose methods are not open-source.
Spatial-AST is evaluated on the binauralized SO-Dataset.
To analyze the effect of data scale, we train and evaluate DCASE baseline encoders on collected open-source data of SO-Dataset (29 classes), and the full SO-Dataset.
We further compare recent stronger open-source SELD systems under the 63-class SO-Dataset protocol: DeepASA \cite{lee2025deepasa}, an object-oriented CNN-Conformer network with an ACCDOA head; PSELDNet \cite{hu2025pseldnets}, SELD networks pretrained on large-scale synthetic data; and GRAM \cite{yuksel2025gram}, a spatial general-purpose audio representation model.
They are trained on SO-Dataset and evaluated under our protocol.

For Spatial Audio LLM evaluation, we compare general LALMs and Omni LLMs, including Qwen2.5-Omni \cite{xu2025qwen25omnitechnicalreport}, Qwen3-Omni \cite{xu2025qwen3}, Phi-4-MM \cite{abouelenin2025phi4mini}, Kimi-Audio \cite{ding2025kimi}, and Audio Flamingo 3 \cite{ghosh2026audio}.
None accepts multichannel input, so each receives a single downmixed channel of the same FOA recording, with identical audio content and only the spatial channels removed.
Since few open-source Spatial Audio LLMs support FOA input, we further design controlled baselines and ablations.
BAT is used as a representative binaural Spatial Audio LLM; FOA audio is converted to binaural input through the FOA-to-binaural interface from SoundSpaces before evaluation.
JAEGER \cite{liu2026jaeger} is the only open-source FOA spatial audio LLM; its official checkpoint outputs only DoA and distance, so we fine-tune it on SO-QA for the full benchmark.

SO-4B, SO-7B, SO-AF3, SO-Kimi, and SO-30B apply the same SO-Encoder, projector interface and training procedure to Phi-4-MM, Qwen2.5-Omni, Audio Flamingo 3, Kimi-Audio, and Qwen3-Omni, respectively.
The variants therefore test whether the spatial upgrade transfers across distinct base-model families rather than only within one Omni LLM.
SO-7B-iv directly feeds basic down-sampled IV features as spatial tokens to evaluate signal-level spatial features.
SO-7B-neuiv follows JAEGER and uses a lightweight CNN and MLP to extract neural IV features.
SO-7B-zs uses zero spatial tokens to verify whether the improvement comes from real spatial token information, and SO-7B-so feeds only spatial tokens without original audio tokens to verify their standalone contribution.
SO-7B-uat is an audio-tower-unfreezing ablation: unlike the default design, it updates the original Qwen2.5-Omni audio tower during spatial adaptation, directly testing whether the frozen semantic-audio pathway is necessary.
Since the original Qwen-Omni training data is unavailable, we mix around 130K open-source monaural QA pairs from Audio Flamingo 3's training data with 20\% of SO-QA at a 1:3 ratio.
The resulting MIX variants use the same recipe across the available LLM backbones; green rows in Table~\ref{tab:results} identify these mixed-data variants.

\section{Results}
\label{sec:results}

\begin{figure}[t]
    \centering
    \includegraphics[width=\columnwidth]{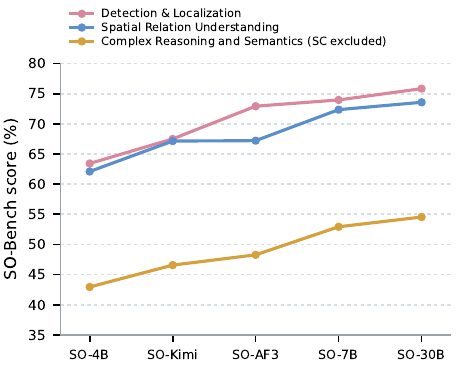}
    \caption{Spatial ability of five Spatial-Omni backbone upgrades. Each point is the mean within one SO-Bench level; Level 3 excludes SC.}
    \label{fig:spatial_scaling}
\end{figure}

\begin{figure*}[t]
    \centering
    \includegraphics[width=0.98\textwidth]{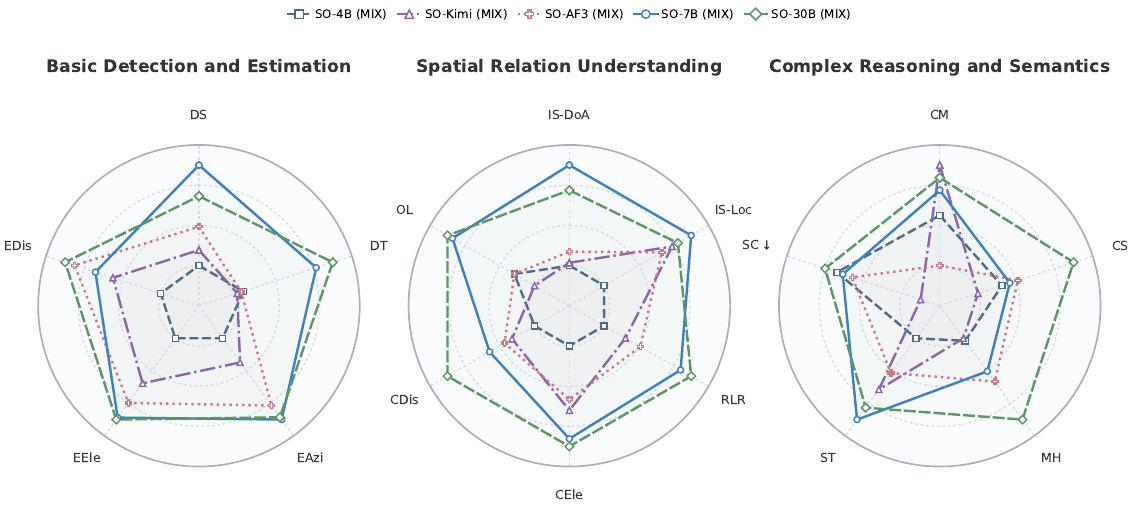}
    \caption{SO-Bench profiles of the five MIX Spatial-Omni upgrades across the three spatial-understanding stages. On SC, the direction is reversed: outward means lower raw WER.}
    \label{fig:spatial_radars}
\end{figure*}
\begin{figure}[t]
    \centering
    \includegraphics[width=\columnwidth]{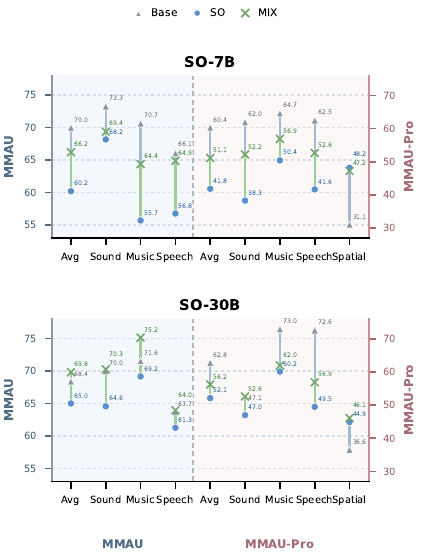}
    \caption{General-audio retention and recovery in absolute scores.  We show Avg, Sound, Music, and Speech for MMAU \& MMAU-Pro, and additionally Spatial for MMAU-Pro. Appendix Table~\ref{tab:mmaus_appendix} provides the full tabular record.}
    \label{fig:mmau_retention}
\end{figure}

We first evaluate SO-Encoder separately, as shown in Table~\ref{tab:encoder_results}.
The DCASE 2024 best \cite{Du_NERCSLIP_task3_report} and second-place \cite{Yu_HYUNDAI_task3a_report} systems are public challenge reports trained on STARSS23 with augmented data and without public code, so they serve as reference upper bounds rather than direct comparisons.
Among reproducible open-source baselines, the DCASE 2024 baseline shows a clear F20 drop at 63 classes. The multi-ACCDOA representation appears limited under many event classes.
Spatial-AST has basic spatial modeling ability on the binauralized SO-Dataset, but it is still weaker than SO-Encoder under the 63-class setting.

\begin{table}[h]
    \centering
    \small
    \setlength{\tabcolsep}{1mm}
    \resizebox{\linewidth}{!}{%
    \begin{tabular}{l|c|ccc}
        \toprule
        Model & Classes & F20(\%) $\uparrow$ & DoA($^\circ$) $\downarrow$ & Rel. Dis. $\downarrow$  \\
        \midrule
        DCASE24 baseline & 13 & 13.1 & 36.9 & 0.33 \\
        \rowcolor{blue!8}
        DCASE24 best & 13 & \textbf{54.4} & {13.6} & \underline{0.21} \\
        \rowcolor{blue!8}
        DCASE24 second & 13 & 29.8 & 19.8 & 0.28 \\
        DCASE24 baseline & 29 & 36.1 & 24.4 & 0.53 \\
        DCASE24 baseline & 63 & 11.2 & 28.1 & 0.33 \\
        Spatial-AST & 63 & 29.2 & 36.0 & 0.36 \\
        DeepASA & 63 & 28.8 & 19.2 & - \\
        PSELDNet & 63 & 48.4 & \underline{12.7} & \underline{0.21} \\
        GRAM & 63 & 25.5 & 24.7 & 0.38 \\
        \midrule
        \textbf{SO-Encoder (Ours)} & 63 & \underline{48.6} & \textbf{7.3} & \textbf{0.20} \\
        \bottomrule
    \end{tabular}}
    \caption{Comparison between spatial audio encoders and existing SELD baselines. Shaded rows indicate closed-source results. DeepASA, PSELDNet, and GRAM are trained or evaluated on SO-Dataset under the same protocol.}
    \label{tab:encoder_results}
\end{table}

SO-Encoder obtains F20=48.6\%, 7.3 degrees DoA error, and 0.20 relative distance error, surpassing all reproducible baselines on every metric.
Among the stronger open-source systems trained on SO-Dataset, PSELDNet benefits from large-scale synthetic pretraining and is the most competitive one at 48.4 F20 and 12.7 degrees DoA error, but its localization remains coarser than ours.
DeepASA reaches a reasonable 19.2 degrees DoA error yet its F20 stays at 28.8, consistent with the limitation of ACCDOA-style outputs under 63 classes, and the general-purpose representation of GRAM transfers poorly to this protocol (25.5 F20, 24.7 degrees).
The DoA error of SO-Encoder is even lower than the closed-source DCASE 2024 best reference, which remains strongest in F20 only on the 13-class STARSS23 setting, so SO-Encoder provides a reliable basis for the spatial tokens consumed by the downstream LLM.

\begin{table*}[t!]
    \centering
    \small
    \setlength{\tabcolsep}{1mm}
    \resizebox{\textwidth}{!}{%
    \begin{tabular}{l|ccccc|cccccc|ccccc}
        \toprule
        \multirow{2}{*}{Model} & \multicolumn{5}{c|}{Basic Detection and Estimation} & \multicolumn{6}{c|}{Spatial Relation Understanding} & \multicolumn{5}{c}{Complex Reasoning and Semantics} \\
        & DS & DT & EAzi & EEle  & EDis & IS-DoA & IS-Loc & RLR & CEle & CDis & OL & CM & CS & MH & ST & SC$\downarrow$ \\
        \midrule
        \multicolumn{17}{c}{\textbf{Open-source Models}} \\
        \midrule
        Qwen-2.5-Omni &6.75 & 55.91 & 10.36 & 32.83 & 56.17 & 34.63 & 24.77 & 49.55 & 50.45 & 59.76 & 37.78 & 22.58 & 29.41 & 10.51 & 11.41 & 85.69\\
        Qwen-3-Omni &13.94 & 54.87 & 11.92 & 28.21 & 43.06 & 21.77 & 39.58 & 49.21 & 51.61 & 46.03 & 35.11 & 17.74 & 26.47 & 0.00 & 9.52 & 89.16\\
        Phi-4-MM &9.75 & 53.33 & 12.31 & 30.39 & 59.05 & 22.34 & 22.22 & 44.74 & 55.86 & \third{64.56} & 49.49 & 20.97 & 23.53 & 1.50 & 7.66 & 84.38\\
        Kimi-Audio &23.84 & 16.07 & 10.81 & 17.99 & 50.00 & 41.08 & 28.53 & 42.64 & 50.75 & 58.56 & 54.21 & 17.74 & 26.47 & 9.01 & 7.17 & 74.48\\
        Audio Flamingo 3 &13.04 & 42.92 & 32.38 & 30.88 & 52.56 & 32.83 & 16.37 & 52.55 & 51.35 & 62.42 & 51.33 & 16.13 & 23.53 & 5.41 & 12.31 & 77.85\\
        \midrule
        \multicolumn{17}{c}{\textbf{Closed-source Models}} \\
        \midrule
        Gemini-2.5-flash &12.14 & 63.15 & 12.16 & 28.04 & 53.91 & 29.39 & 13.21 & 42.34 & 49.25 & 55.56 & 43.33 & 22.58 & 38.24 & 3.30 & 8.31 & 88.85\\
        Gemini-2.5-pro &14.24 & 72.72 & 11.71 & 31.18 & 36.42 & 34.33 & 21.47 & 50.75 & 49.25 & 58.86 & 33.06 & 24.19 & 38.24 & 4.50 & 4.71 & 81.19\\
        Gemini-3-pro &16.34 & 58.90 & 12.76 & 30.13 & 58.97 & 36.28 & 28.98 & 38.64 & 53.49 & \underline{68.18} & 45.00 & 35.90 & 35.29 & 18.60 & 9.21 & \underline{73.46}\\
        GPT-audio &3.03 & 36.15 & 4.48 & 24.24 & 28.21 & 26.87 & 11.94 & 64.52 & 45.16 & 46.88 & 5.00 & 17.95 & 23.53 & 0.00 & 4.64 & 96.10\\
        \midrule
        \multicolumn{17}{c}{\textbf{Spatial Baselines}} \\
        \midrule
        SO-7B-iv &45.28 & 82.67 & 17.27 & 30.73 & 63.79 & 56.37 & 54.35 & 49.85 & 47.75 & 60.06 & 82.55 & 43.55 & \third{50.00} & 22.52 & 25.38 & 76.03\\
        SO-7B-neuiv &44.68 & 84.53 & 31.83 & 48.73 & 70.58 & 55.77 & 54.05 & 62.16 & 52.25 & 61.86 & 81.25 & 41.94 & 23.53 & 24.32 & 33.60 & 76.98\\
        SO-7B-zs &21.10 & 71.21 & 13.06 & 31.18 & 48.56 & 51.72 & 40.69 & 47.15 & 49.55 & 50.15 & 53.58 & 38.24 & 20.59 & 22.52 & 10.21 & 77.11\\
        SO-7B-so &35.23 & 53.18 & 62.31 & 72.41 & 82.51 & 41.83 & 39.49 & 51.65 & 45.05 & 48.65 & 40.25 & 14.52 & 20.59 & 15.62 & 6.61 & 98.73\\
        SO-7B-uat &\textbf{52.02} & 84.10 & 70.22 & 76.75 & \underline{85.39} & 64.61 & 58.70 & 48.64 & 64.56 & 61.86 & 83.16 & 64.51 & 29.41 & 28.82 & 17.41 & \third{73.49}\\
        BAT &30.97 & 58.19 & 52.10 & 47.67 & 53.91 & 62.67 & 58.56 & 52.85 & 52.25 & 52.85 & 72.90 & 43.55 & 32.35 & 25.53 & 16.87 & 98.40\\
        JAEGER &46.48 & 82.04 & 45.31 & 52.21 & 68.11 & 57.12 & 54.65 & 54.35 & 51.35 & 50.15 & 81.31 & 59.68 & 38.24 & 24.32 & 35.96 & 78.58\\
        \midrule
        \multicolumn{17}{c}{\textbf{Spatial-Omni Models}} \\
        \midrule
        SO-4B &50.07 & 79.04 & 56.51 & 55.59 & 75.93 & 62.97 & 60.21 & 62.46 & 54.95 & 53.15 & 78.85 & 62.90 & 47.06 & 35.44 & 26.43 & 75.92\\
        \rowcolor{green!10}
        SO-4B (MIX) &49.18 & 79.15 & 55.30 & 53.08 & 73.46 & 63.42 & 59.91 & 63.66 & 56.46 & 51.95 & 79.47 & 66.13 & 41.18 & 33.93 & 33.33 & 77.06\\
        SO-AF3 &50.22 & 79.09 & 75.95 & 75.02 & 84.36 & 63.72 & 65.32 & 67.87 & 70.57 & 57.36 & 78.44 & 62.90 & 47.06 & 39.34 & 43.83 & 79.70\\
        \rowcolor{green!10}
        SO-AF3 (MIX) &49.93 & 78.93 & 75.61 & 75.63 & \third{84.57} & 64.17 & 65.77 & 68.77 & 71.17 & 57.96 & 79.47 & 59.68 & 47.06 & 38.74 & 43.84 & 80.84\\
        SO-Kimi &49.62 & 78.16 & 63.02 & 66.86 & 79.83 & 63.56 & 65.31 & 66.06 & 74.17 & 55.55 & 78.23 & \third{69.35} & 32.35 & 35.43 & 49.18 & 97.70\\
        \rowcolor{green!10}
        SO-Kimi (MIX) &49.48 & 78.47 & 62.59 & 68.78 & 79.63 & 63.57 & \third{66.82} & 66.67 & 74.17 & 56.46 & 77.00 & \textbf{72.58} & 32.35 & 33.63 & 48.83 & 97.54\\
        SO-7B &\underline{51.72} & 87.01 & 76.38 & 76.92 & 77.77 & \underline{68.06} & 66.81 & 72.97 & \third{81.68} & 57.95 & \third{86.65} & 67.74 & 47.05 & \third{39.93} & \underline{57.02} & 75.94\\
        \rowcolor{green!10}
        SO-7B (MIX) &\third{51.12} & \third{87.15} & \textbf{79.86} & \underline{80.74} & 81.89 & \textbf{68.81} & \textbf{68.76} & \third{74.47} & \underline{81.98} & 60.96 & \underline{87.06} & \third{69.35} & 44.11 & 37.53 & \textbf{57.94} & 78.49\\
        SO-30B &50.82 & \underline{87.39} & \third{76.82} & \third{79.79} & 84.36 & 67.17 & 64.41 & \textbf{76.88} & \underline{81.98} & \third{64.56} & 86.45 & 59.68 & \underline{61.76} & \underline{43.23} & 53.53 & \textbf{71.65}\\
        \rowcolor{green!10}
        SO-30B (MIX) &50.52 & \textbf{88.97} & \underline{79.16} & \textbf{81.43} & \textbf{85.80} & \third{67.46} & \underline{67.41} & \underline{75.97} & \textbf{84.08} & \textbf{69.39} & \textbf{87.67} & \underline{70.96} & \textbf{67.64} & \textbf{43.24} & \third{54.34} & 74.18\\
        \bottomrule
    \end{tabular}}
    \caption{Main results of general LALMs, Omni LLMs, spatial baselines, and Spatial-Omni variants. Green rows are mixed-data variants. Bold, underline, and blue indicate the best, second, and third result in each column. Metric abbreviations are listed in Section~\ref{sec:benchmark}; all metrics are higher-is-better except SC (WER), for which lower is better.}
    \label{tab:results}
\end{table*}

Table~\ref{tab:results} reports the main SO-Bench results.
General LALMs and Omni LLMs are not helpless on basic audio and coarse reasoning: some Gemini models do reasonably on DT, CDis, and CS, and reasonable spatial position comparison abilities, which may reflect semantic priors, loudness cues, or answer-format biases in simple comparisons.
On explicit spatial estimation and grounded reasoning (EAzi, EEle, IS-Loc, MH, and ST), they remain weak, so monaural input with general audio-language training does not support reliable spatial audio understanding by itself.

BAT, an existing binaural Spatial Audio LLM, performs well on IS-DoA, IS-Loc, and CM, matching the benefit of explicit spatial training for source location and motion.
Trained mainly on binaural two-source simulated data with clip-level prediction, it falls behind on precise angle estimation, complex multi-source relations, and speech content recognition.
The official JAEGER checkpoint supports only DoA and distance outputs and performs poorly on the three comparable subtasks.
After fine-tuning on SO-QA, JAEGER becomes the strongest spatial baseline and leads on CM and ST, yet it still trails Spatial-Omni models on precise angle and distance estimation and source identification; because fine-tuning aligns prompts and data, the remaining gap is unlikely to be a prompt mismatch alone, although the architectures and base LLMs also differ.

The Spatial-Omni family is strongest across the benchmark.
As shown in Table~\ref{tab:results}, SO variants achieve best and second-best scores on 14 of 16 tasks, and the remaining three are close to the best.
SO-7B(MIX) reaches 79.86 on EAzi, 80.74 on EEle, 68.76 on IS-Loc, and 74.47 on RLR, while SO-30B(MIX) reaches 88.97 on DT, 84.08 on CEle, 69.39 on CDis, and 70.96 on CM.
Comparing to base spatial only training, mixed training can retain strong spatial performance rather than merely recover general-audio behavior.
The direct SO-7B and SO-30B variants remain competitive on complementary tasks, including SO-7B's 81.68 CEle and SO-30B's 81.98 CEle and 61.76 MH.
SO-4B row confirms that the interface also upgrades smaller models; its MIX version is reported alongside it rather than being inferred from incomplete evaluation.
SO-AF3 and SO-Kimi extend the upgrade to LALMs and prove that without original omni ablities, the extra spatial token interface can still provide strong spatial reasoning and estimation.

Some tasks remain challenging: the overall  CS score is low, and the multi-hop reasoningscore is below 50 for all models.
SO variants also do not show large gap on CDis task: binary comparisons can be answered from coarse loudness or response priors, whereas SO-7B leans on spatial-token evidence, which pays off in precise estimation more than in two-choice comparisons.

The controlled baselines separate the contributions of different spatial features.
Raw IV features carry spatial energy and directional-motion cues, so SO-7B-iv performs well on DS, DT, OL, CM, and CS, but basic signal features run out of expressiveness on precise angle estimation, multi-hop reasoning, and spatial-temporal description.
SO-7B-neuiv improves over direct IV input and better score, so a lightweight neural extractor organizes IV features better for temporal localization.
SO-7B-so(spatial only), which feeds spatial tokens without the original audio semantic tokens, keeps strong EAzi and EEle but drops clearly on DS, IS-DoA, IS-Loc, CM, ST, and SC.
SO-Encoder alone thus supplies accurate geometry, but the LLM still needs semantic audio tokens to identify events and bind them to locations.
SO-7B-zs(zero spatial) retains part of the time localization and coarse spatial ability, so the interface itself is harmless; its large gap to models with real spatial features attributes the main improvement to informative spatial tokens rather than to the interface.
The SO-7B-uat(unfreeze audio tower) ablation shows why we retain the default frozen audio tower. Although unfreezing raises a few individual scores (for example EDis), it does not give a uniform gain: RLR drops from 72.97 to 48.64 and MH from 39.93 to 28.82 relative to the default SO-7B. This trade-off supports preserving the base semantic-audio pathway during spatial adaptation.

We provide extra benchmark results on general audio understanding using MMAU and MMAU-Pro \cite{sakshi2025mmau,kumar2026mmau} in Figure~\ref{fig:mmau_retention}.
Figure~\ref{fig:mmau_retention} jointly summarizes the absolute MMAU and MMAU-Pro scores for the 7B and 30B backbones.
For the 7B backbone, spatial adaptation changes the MMAU and MMAU-Pro averages by $-9.79$ and $-18.53$ points, respectively, while MIX recovers $+6.00$ and $+9.32$ points relative to the spatially adapted model.
The MMAU-Pro spatial category instead gains $+17.09$ points after spatial adaptation and remains $+16.17$ points above the base model after MIX, indicating that the recovery does not erase the spatial benefit.
For the 30B backbone, the corresponding average changes are $-3.40$ on MMAU and $-10.71$ on MMAU-Pro; its MIX variants recover $+4.80$ and $+4.10$ points from the SO models.
On MMAU-Pro Spatial, SO-30B improves by $+8.30$ points and the full-length MIX run adds a further $+1.23$ points.
Although the MMAU-Pro spatial category is a binaural benchmark, we convert it into pseudo-FOA format and gain better results after MIX.
Together, these trajectories show that mixed training can restore general-audio performance while preserving and even promoting the spatial-audio gain.
Further results show that pure upgrade to base LLM or LALMs is easier than to Omni LLMs. Spatial SFT may cause more general-audio degradation on Omni LLMs, but MIX can still recover the performance more easily than on base LLMs. 
More MIX results are provided in Appendix Table~\ref{tab:mmaus_appendix}.
\section{Conclusion and Discussion}
\label{sec:conclusion}

We propose Spatial-Omni to improve spatial audio understanding in LALMs and Omni LLMs.
Its SO-Encoder is a lightweight parallel spatial encoding branch that, without modifying the original audio encoder, extracts direction, distance, motion, and multi-source relation cues from FOA audio and maps them into spatial tokens for the LLM.
Trained with SO-Dataset, SO-QA, and benchmark SO-Bench, which we build from open-source data, real recordings, and simulated scenes, Spatial-Omni models outperform general LALMs, Omni LLMs, and existing spatial audio baselines on most SO-Bench tasks.
General audio ability still degrades without the original training data, but is partially recovered by mixed training.

\section*{Limitations}
Spatial-Omni improves spatial audio understanding, but several limitations remain.
First, our current study focuses on FOA input under a unified coordinate convention.
Although FOA is widely used in spatial audio research, the model has not been systematically evaluated on other spatial representations or microphone geometries, such as SALSA and SALSA-Lite.

Second, SO-Encoder relies on track-level supervision, matching, and activation thresholds inherited from SELD-style training.
These design choices may limit source counting and relation reasoning in scenes with many simultaneous sources, short events, or ambiguous spatial overlaps.

Third, our model currently supports audio inputs of up to 20 s. Further exploration may include longer audio inputs.

Finally, Spatial-Omni is designed to preserve the base Omni LLM's semantic audio pathway, but some degradation on general audio benchmarks remains.
This suggests that spatial adaptation and general audio capability still need better balancing, for example through broader monaural-spatial mixed instruction tuning.

\bibliography{custom}

\newpage
\appendix

\section{Related Work Details}
\label{app:relatedwork_details}

\subsection{Spatial Audio LLMs}
Existing Spatial Audio LLMs extend spatial audio understanding into the LLM framework, usually by designing encoders that can accept spatial audio inputs and answer spatial questions.
One line of work focuses on binaural spatial audio.
BAT~\cite{zheng2024bat} designs Spatial AST, an audioMAE-based spatial audio encoder.
It accepts binaural input and learns spatial representations from simulated data.
Hear You Are~\cite{anonymous2026hear} adds visual input on top of the BAT spatial audio encoder and evaluates spatial audio understanding in 360 degrees video tasks.
SING~\cite{mishra2025spatial} adds a DoA encoder so that a large model can recognize speaker content from a specified direction.
OWL~\cite{biswas2025owl} supports binaural spatial audio and explicitly models RIRs to learn spatial information for occluded scene reasoning.
These methods show that explicit spatial cues are useful for LLM-based acoustic scene understanding.
However, they mainly rely on binaural formats.
Much of their training and evaluation data comes from simulation.

Another line of work focuses on FOA or multichannel input.
~\cite{tang2024can} introduces LLM understanding for FOA spatial audio and injects FOA IV vectors into the base audio encoder so that the model can learn spatial information from spatial audio.
SPUR~\cite{sakshi2025spur} modifies the original audio encoder into a Spatial Encoder for FOA input.
It uses 3D convolution to extract spectral-spatial covariance features for stronger representation in multi-source scenes.
Sci Phi~\cite{jiang2026sci} trains a Spatial Encoder on a large simulated dataset and uses it in parallel with an Audio Encoder to improve spatial understanding.
JAEGER~\cite{liu2026jaeger} uses neural IV features to help models learn speaker direction in 3D understanding.
PhaseCoder~\cite{dementyev2026phasecoder} treats different microphone arrangements as an input modality, allowing LLMs to learn directional representations across spatial audio formats and better adapt to microphone geometry changes.
The World is Not Mono~\cite{you2026world} introduces an expert system that fuses semantic encoder and spatial encoder outputs before feeding them into the LLM, reducing the mono bias of pretrained models.
ST-Omni~\cite{zeng2026listen} follows BAT and further explores spatial audio in world model scenes.
These studies show that spatial audio can enhance LLM spatial understanding, but many methods still bind spatial modeling to modification or retraining of the original audio encoder.

\subsection{Spatial Audio Benchmarks}
\begin{table*}[t]
\centering
\small
\resizebox{\textwidth}{!}{
\begin{tabular}{l|c|c|c|c|ccc}
\toprule
\multirow{2}{*}{\textbf{Benchmark}} & \multirow{2}{*}{\textbf{Modality}} & \multirow{2}{*}{\textbf{Format}} & \multirow{2}{*}{\textbf{Size (QA/Audio)}} & \multirow{2}{*}{\textbf{Source}} & \multicolumn{3}{c}{\textbf{Task Support}} \\
 & & & & & \textbf{Loc} & \textbf{Reason} & \textbf{Motion} \\
\midrule
BAT Dataset & A & Binaural  & 872K QA & Simulated & \checkmark & \checkmark & $\times$ \\
AudioMotionBench & A & Binaural & 224 clips / 1K QA & Simulated & $\times$ & \checkmark & \checkmark \\
STAR Bench & A & Binaural & 2K QA & Simulated & \checkmark & \checkmark & \checkmark \\
Hear You Are QA & A+V & Binaural  & 1M QA & Simulated & \checkmark & \checkmark & $\times$ \\
SAVVY Bench & A+V & 7 ch & 1.5K QA & Recorded & \checkmark & \checkmark & \checkmark \\
SPUR Set & A & FOA  & 18k QA & Mixed & \checkmark & \checkmark & \textbf{Partial} \\
BiDepth (OWL) & A+V & Binaural  & 28k clips / 1.1M QA & Simulated & \checkmark & \checkmark & $\times$ \\
The World is Not Mono & A & Binaural & unknown & Simulated & \checkmark & \checkmark & \checkmark \\
Sci Phi & A & FOA  & 1.6 M clips & Simulated & \checkmark & \textbf{Partial} & \textbf{Partial} \\
MMAU-Pro & A & Binaural & 325 QA & Recorded & \checkmark & \checkmark & \checkmark \\
\midrule
\textbf{SO-Dataset (Ours)} & A+V & FOA & 400K clips / 2.1M QA & Mixed & \checkmark & \checkmark & \checkmark \\
\textbf{SO-Bench (Ours)} & A+V & FOA & 7K clips / 7K QA & Mixed & \checkmark & \checkmark & \checkmark \\
\bottomrule
\end{tabular}}
\caption{Comprehensive comparison of spatial audio datasets and benchmarks.}
\label{tab: spatial_audio_benchmarks}
\end{table*}
Existing benchmarks leave gaps for evaluating FOA spatial audio reasoning.
General MLLM benchmarks mainly evaluate semantic understanding ~\cite{yu2023mm, yu2024mm, yue2024mmmu, liu2024mmbench}, while visual spatial benchmarks lack systematic spatial audio evaluation ~\cite{azuma2022scanqa, xu2025spatialbench}.
Recent audio benchmarks cover binaural motion, audio-visual viewpoint reasoning, or partial spatial tasks ~\cite{sun2025spatial, sridhar2026spatial, chen2026savvy, liu2025star, kumar2026mmau}.
They remain limited for FOA audio, multiple-source relations, motion analysis, and comprehensive spatial reasoning.
SO-Bench is designed for FOA spatial audio and covers localization, relative position understanding, motion analysis, and complex spatial question answering.
Table~\ref{tab: spatial_audio_benchmarks} summarizes representative spatial audio datasets and benchmarks.

\section{Details of Dataset}
\label{app: dataset}

\subsection{Collected Open-Source Datasets}
We use the following open-source datasets for training: L3DAS22 and L3DAS23~\cite{guizzo2022l3das22, marinoni2023overview}, TAU Spatial Sound Events 2019, 2020, 2021~\cite{adavanne2019tau, adavanne2019moving, politis2020taunigens, politis2021taunigens}, and STARSS22, STARSS23~\cite{politis2022starss22,shimada2023starss23}.
The STARSS datasets include additional visual information. For real data with missing distance and elevation information, we filter out the corresponding losses during encoder training to avoid affecting azimuth and sound event recognition training.
The distribution of these datasets is summarized in Table~\ref{tab:open_source_dataset}.

\begin{table}[htb]
    \centering
    \small
    \begin{tabular}{l|c}
        \toprule
        \textbf{Dataset} & \textbf{Clips} \\
        \midrule
        TAU 2019 Moving & 10.6k \\
        TAU 2019 & 1.3k \\
        TAU 2020 & 1.8k \\
        TAU 2021 & 1.8k \\
        L3DAS 22 & 2.3k \\
        L3DAS 23 & 7.7k \\
        STARSS 22 & 0.9k \\
        STARSS 23 & 1.4k \\
        \bottomrule
    \end{tabular}
    \vspace{0.2cm}
    \caption{Distribution of open-source datasets.}
    \label{tab:open_source_dataset}
\end{table}

\subsection{Recorded Real World Dataset}
The recorded subset complements open-source SELD data and simulated data with real acoustic conditions and synchronized visual context, following the practice of real-scene audiovisual SELD datasets such as STARSS22 and STARSS23~\cite{politis2022starss22,shimada2023starss23}.
It contains 3.5k FOA clips from 15 scene types, 23 recording scenes, covering both indoor and outdoor environments.
The scenes include offices, kitchens, seminar rooms, corridors, campus roads, public activity areas, gates, and sports fields.
These recordings introduce natural reverberation, background noise, occlusion, moving sources, and visually observable sound events that are difficult to fully reproduce with simulation.
Table~\ref{tab:recorded_scene_statistics} summarizes the raw recording sessions before annotation-based segmentation.

\begin{table*}[t]
    \centering
    \small
    \setlength{\tabcolsep}{4pt}
    \renewcommand{\arraystretch}{1.08}
    \begin{tabularx}{\textwidth}{l | X |c}
        \toprule
        \textbf{Scene Type} & \textbf{Representative Content} & \textbf{Duration} \\
        \midrule
        \multicolumn{3}{c}{\textbf{Indoor}} \\
        \midrule
        Kitchen & Cooking, tableware sounds, object impacts, and speech. & 120 min \\
        Supermarket & Checkout, announcements, carts, and customer speech. & 70 min \\
        Laboratory/office & Conversations, typing, coffee machines, and printer events. & 60 min \\
        Dormitory common area & Walking, conversations, key sounds, and door access events. & 30 min \\
        Study corridor & Quiet ambience, footsteps, passing vehicles, and coffee machines. & 30 min \\
        Parcel station & Package packing, barcode scanning, checkout sounds, and speech. & 20 min \\
        Print shop & Multiple visible and off-screen printer events. & 15 min \\
        Seminar room & Discussion, chair movement, and whiteboard interaction. & 15 min \\
        Tennis court & Tennis hits and ball-bounce events. & 15 min \\
        Table tennis court & Table-tennis hits and ball-bounce events. & 8 min \\
        \midrule
        \multicolumn{3}{c}{\textbf{Outdoor}} \\
        \midrule
        Entrance gate & Turnstiles, passing pedestrians, vehicles, and machine prompts. & 70 min \\
        Escalator/metro entrance & Escalators, warning prompts, broadcasts, footsteps, and speech. & 40 min \\
        Subway platform & Train arrivals/departures, broadcasts, and warning tones. & 30 min \\
        Campus road & Pedestrians, conversations, rolling luggage, vehicles, and whistles. & 20 min \\
        Public square & Outdoor ambience, speech, vehicles, unpacking, and music. & 10 min \\
        \bottomrule
    \end{tabularx}
    \caption{Statistics of raw real-world recording sessions used to construct the recorded subset. Durations are measured before event-level annotation and clip segmentation.}
    \label{tab:recorded_scene_statistics}
\end{table*}

\paragraph{Scene}
We plan the recorded subset around two types of real environments: indoor spaces and public open spaces.
Indoor recordings focus on offices, kitchens, seminar rooms, and corridors, where local spatial relations, near-field reflections, speech, door movement, object impacts, typing, appliances, and kitchen activities can be observed clearly.
Public open-space recordings focus on campus roads, activity areas, gates, and sports fields, where moving sources, far-field propagation, traffic-like sounds, and less constrained background noise are more common.
When selecting a recording position, we prefer open viewpoints with clear entering and leaving paths for sound sources, and avoid locations with severe crowd occlusion when possible.
We also intentionally include scenes where target events occur frequently, so that the recorded data contains enough useful event activity rather than long silent or irrelevant background segments.

\paragraph{Recording}
We use an Insta360 Pro camera to record the 360-degree panoramic visual scene and a ZOOM H3-VR Ambisonic microphone to capture FOA spatial audio.
The microphone is fixed close to the panoramic camera, with the acoustic center placed near the visual center to reduce spatial offset between modalities.
The recordings are collected in two modes.
For controlled events that require explicit event triggering or performer cooperation, participants are recruited to take part in data collection, and all participants sign informed consent forms before recording.
To protect privacy, we avoid close-up facial capture when possible by keeping the panoramic camera away from participants' faces and conservatively screening recordings that may contain identifiable personal information.
For public scenes, we obtain permission from the relevant venue managers before recording natural sound events.
During each session, we record scene layout, important sound events, approximate source positions, source movement patterns, and recording context.
These field notes provide reference information for later synchronization, annotation, and quality checking.

\paragraph{Synchronization}
The raw recordings are first coarsely synchronized using the start-time offset between the video and external FOA audio.
We then refine the alignment by comparing the audio track embedded in the panoramic video with the external FOA recording.
Transient events, clear onsets, and rhythmic sound boundaries are used as references for frame-level manual correction.
This step produces aligned long recordings, where the panoramic video and the external FOA audio share the same timeline before annotation and segmentation.

\paragraph{Annotation}
After audio-video synchronization, annotators use the video stream and the binaural audio embedded in the panoramic video as the main reference for annotation.
This embedded audio is easier to audition together with the visual scene and is tightly bound to the video timeline, which reduces boundary errors caused by small offsets between the external FOA signal and the video.
All event-level annotations are completed on the aligned long recordings before clip segmentation.
For each event, we annotate event category, active time interval, track ID, azimuth, elevation, and distance in the DCASE-style coordinate system used in recent SELD-with-distance benchmarks~\cite{aparicio2024baseline} when reliable spatial evidence is available.
When a source can be visually and acoustically localized, we record its relative position and convert it into direction and distance annotations.
When precise 3D localization is unreliable because of occlusion, distance, weak visual evidence, or strong reverberation, we keep coarser direction information instead of forcing an uncertain position label.
For dynamic sources, we record motion-related metadata and generate frame-level spatial labels by interpolating between annotated source states.
This annotation procedure keeps temporal, semantic, and spatial labels in a unified event-level format.

We hire annotators with prior experience in audio annotation and provide them with training on spatial audio concepts, annotation tools, and quality standards. They are paid at an hourly rate of \$40, yielding a total experimental cost of approximately \$2500. Prior to participation, subjects are informed that their assessments will be utilized exclusively for academic research purposes only.

\paragraph{Quality control}
We check the consistency between video, FOA audio, event boundaries, and spatial annotations during post-processing.
Segments with failed synchronization, unclear event identity, corrupted audio, or unreliable spatial evidence are removed or downgraded to weaker supervision.
The final recorded subset provides realistic spatial examples paired with 360-degree visual data, and is used to improve robustness on real scenes and to support multimodal spatial understanding tasks.

\subsection{Simulated Dataset}
We use SoundSpace 2.0~\cite{chen2022soundspaces} to simulate FOA data. The room data is sourced from the HM3D, MP3D~\cite{ramakrishnan2021habitat}, and Replica~\cite{straub2019replica} datasets, totaling 207 rooms. The specific train/valid/test distribution is shown in Table~\ref{tab:room_distribution}.
\begin{table}[htb]
    \centering
    \begin{tabular}{c|c|c|c}
        \toprule
        \textbf{Dataset} & \textbf{Train} & \textbf{Valid} & \textbf{Test} \\
        \midrule
        HM3D & 80 & 10 & 10 \\
        MP3D & 79 & 4 & 6 \\
        Replica & 14 & 2 & 2 \\
        \bottomrule
    \end{tabular}
    \vspace{0.2cm}
    \caption{Distribution of rooms in the simulated dataset.}
    \label{tab:room_distribution}
\end{table}

We also measure the RT60 data of the rooms, as shown in Table~\ref{tab:rt60_distribution}.

\begin{table}[htb]
    \centering
    \begin{tabular}{c|c|c|c|c}
        \toprule
        \textbf{Dataset} & \textbf{Mean} & \textbf{P25} & \textbf{P50} & \textbf{P75} \\
        \midrule
        HM3D & 0.248 & 0.230 & 0.246 & 0.260 \\
        MP3D & 0.251 & 0.153 & 0.194 & 0.268 \\
        Replica & 0.210 & 0.169 & 0.197 & 0.262 \\
        \bottomrule
    \end{tabular}
    \vspace{0.2cm}
    \caption{RT60 distribution of rooms in the simulated dataset.}
    \label{tab:rt60_distribution}
\end{table}

\subsection{Dataset Annotation}
All data is annotated using the DCASE format coordinate system, with CSV files recording time frames at a 10 Hz sampling rate, track, sound event, azimuth, elevation, and distance.
The DCASE coordinate system is used, where +x is forward, +y is left, and +z is up. Azimuth increases counterclockwise and ranges from [-180, 180], elevation increases upward and ranges from [-90, 90], and distance is measured in centimeters.

\begin{figure*}[htb]
    \centering
    \includegraphics[width=0.8\textwidth]{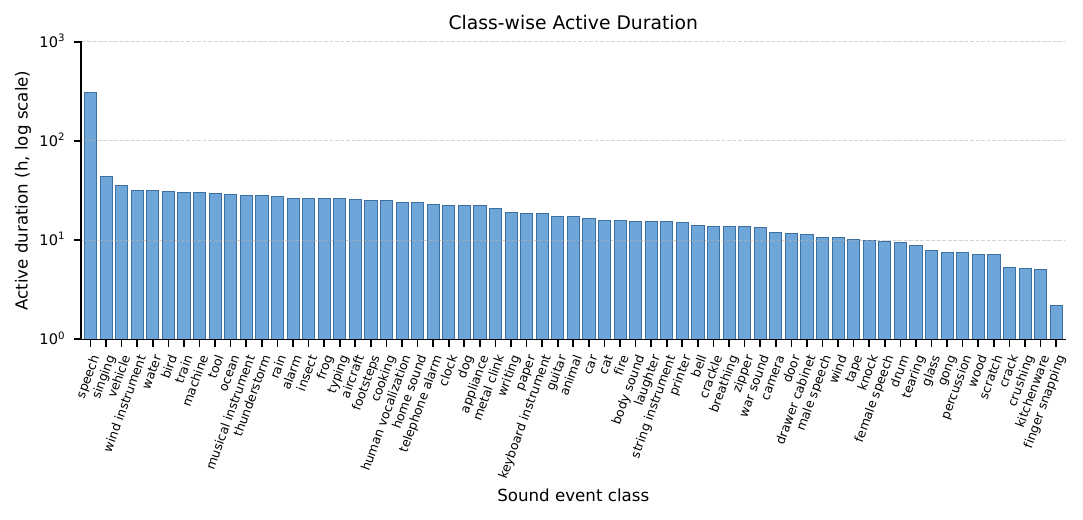}
    \caption{Distribution of sound events in the dataset.}
    \label{fig: sound_event_distribution}
\end{figure*}

For sound label, we aggregate semantically similar labels from FSD50K to form our final sound event label system. The distribution of the 63 dry sound events is shown in Figure~\ref{fig: sound_event_distribution}.

FOA channels are processed in the AmbiX/ACN order: $[ W, Y, Z, X ]$.

\begin{figure*}
    \centering
    \includegraphics[width=0.75\textwidth]{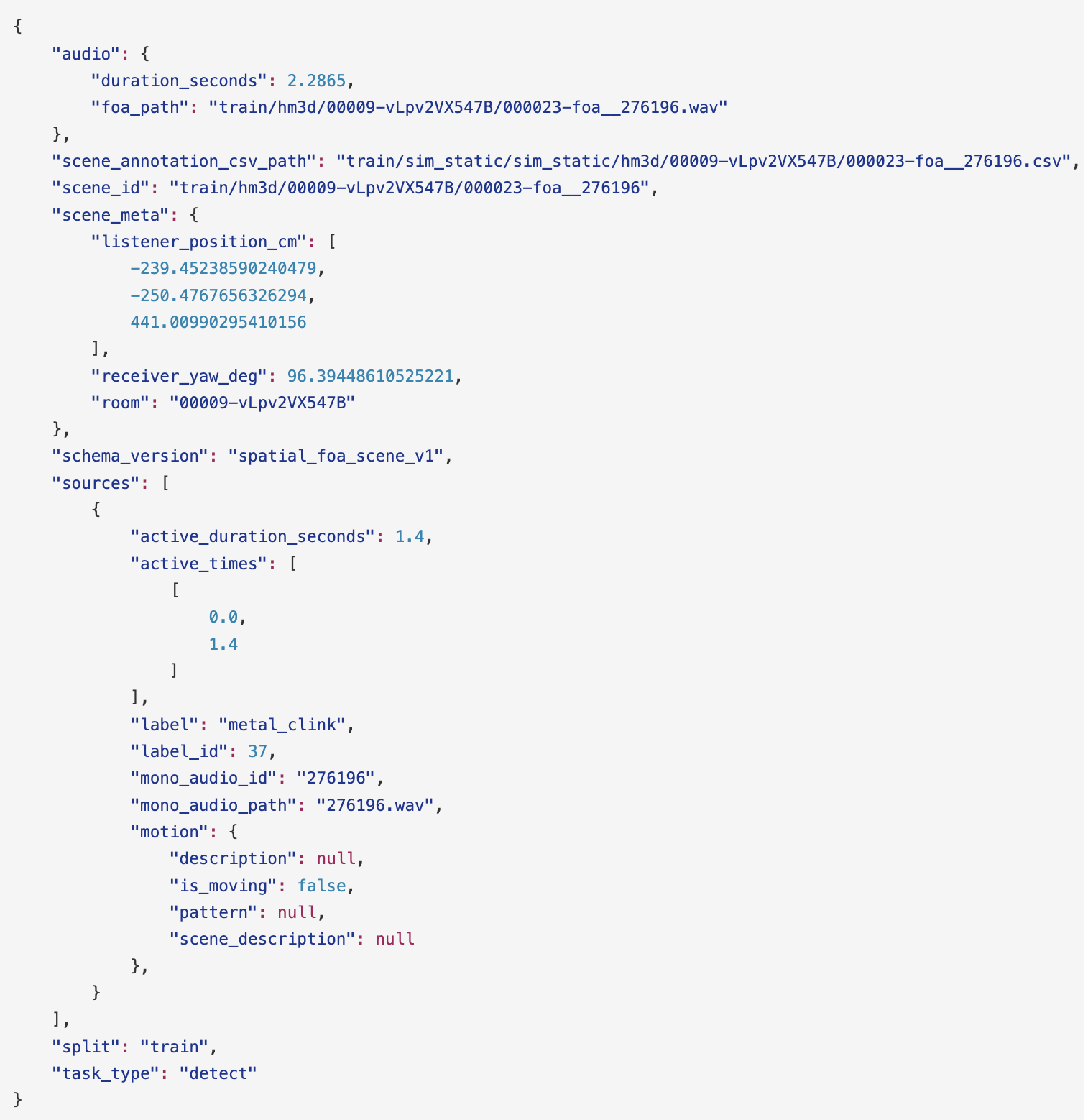}
    \caption{Example of annotation JSON file.}
    \label{fig: annotation_json_example}
\end{figure*}

An example annotation JSON file is shown in Figure~\ref{fig: annotation_json_example}. 
JSON files record the listener's position and orientation, while CSV files record the sound source's relative position to the listener. 
They also include basic sound event labels, active time of sound events, whether they are dynamic or static, and other information.

\subsection{QA Annotation}
For QA annotation, we use a template-guided generation and verification process.
For each spatial audio subtask, we manually design 20--25 question templates and answer templates.
We then use Gemini-3 to instantiate questions with different phrasings from the audio and metadata, including sound events, directions, distances, and motion information in the annotation JSON.
To improve language diversity and reduce template bias, we use GPT-4o to paraphrase all QA pairs.
After annotation, human annotators randomly check a portion of the QA pairs for quality control.
All QA annotation in SO-Bench is checked manually to ensure the quality of the QA pairs.
Examples of QA pairs and prompts are shown in Table~\ref{tab:qa_example}.
\begin{table*}
    \centering
    \small
    \begin{tabular}{c|>{\raggedright\arraybackslash}p{0.78\textwidth}}
        \toprule
        \textbf{Tasks} & \multicolumn{1}{c}{\textbf{QA Pair}} \\
        \midrule
        \multirow{2}{*}{Detect Source} & Q: "Listen to the audio clip and answer based only on what you hear. Which sound source is located to the back-right and below?", \\ & A:  "The sound source located to the back-right and below is breathing." \\
        \midrule
        \multirow{2}{*}{Estimate Azimuth} & Q: "Listen to the audio clip and answer based only on what you hear. Use the DCASE FOA coordinate system: +x front, +y left, +z up; azimuth is in [-180, 180] degrees with positive values to the left, and elevation is in [-90, 90] degrees with positive values upward. From which azimuth is the footstep coming? Report the final angle in degrees. ", \\ & A:  "99.7 degrees" \\
        \midrule
        \multirow{2}{*}{Compare Distance} & Q: "Listen to the audio clip and answer based only on what you hear. Between the sound of frog and the guitar, which sound source is positioned closer to the listener?", \\ & A:  "The guitar is closer to the listener than the frog." \\
        \midrule
        \multirow{2}{*}{Multi-Hop} & Q: "Listen to the audio clip and answer based only on what you hear. Among all detected sound sources, which one is farthest to the left?" \\ & A:"The sound source farthest to the left is the voice of singing." \\
        \midrule
        \multirow{2}{*}{Spatial Temporal} & Q: "Listen to the audio clip and answer based only on what you hear. At what time does the speech sound originating from behind the listener begin?" \\
        & A: "The sound coming from the back becomes active at 0.2s." \\
        \midrule
        \multirow{2}{*}{Speech Content} & Q: "Listen to the audio clip and answer based only on what you hear. What is the speech content originating from the back-right position? Answer in one sentence by transcribing the spoken content as accurately as possible."\\
        & A: "The voice coming from the back-right says, 'A voice inquired. Who's there.'" \\
        \bottomrule
    \end{tabular}
    \vspace{0.2cm}
    \caption{Example of prompts and QA pairs for different tasks.}
    \label{tab:qa_example}
\end{table*}

\subsection{Data Split}
\label{app: data_split}
We strictly ensure the unseen nature of the test set in our data split. For dry sound data, we separately extract unseen sound event mono audio from the eval set of FSD50k. The RIRs in the test set are all from unseen rooms and convolved with dry audio from the eval set. For collected datasets, we follow the split of open-source datasets and construct QA pairs from the test set. For recorded data, we split by recording scenes to ensure that the test set includes unseen scenes and spatial relationships.

We split the training and test sets as follows: for audio clips, the training set contains about 363k clips and the validation set contains 35k clips. 
The train/valid ratio is kept across different data sources to ensure a balanced representation of real and simulated data in both sets.
For the QA set, the training set contains around 2.0M QA pairs and the validation set contains around 100k QA pairs. The test set consists of the 7k QA pairs in the benchmark.

\section{Training Details}
\label{app: training_details}

\subsection{Audio Preprocess}
\label{app: audio_preprocess}
For audio preprocessing, we input $16\,\mathrm{kHz}$ FOA audio with a maximum clip length of $20\,\mathrm{s}$. We use the same STFT parameters as the original Qwen training, allowing us to directly input the mel features extracted from the W channel into the original audio encoder for feature extraction without additional adapter layers.
The details of the parameter configuration are shown in Table~\ref{tab:v13d-foa-input}.

\begin{table*}[h]
\centering
\small
\begin{tabularx}{\textwidth}{@{}l>{\raggedright\arraybackslash}p{0.28\textwidth}X@{}}
\toprule
\textbf{Parameter} & \textbf{Value} & \textbf{Notes} \\
\midrule
Sample rate & $16\,\mathrm{kHz}$ & $4$-channel FOA, order $[W,Y,Z,X]$ \\
Clip duration & $10\,\mathrm{s}$ & Waveform shape $[B,4,160000]$ \\
STFT $n_{\text{fft}}$ & $400$ & Aligned with Qwen2.5-Omni \\
STFT hop length & $160$ & $10\,\mathrm{ms}$ hop \\
STFT window length & $400$ & $25\,\mathrm{ms}$ window \\
Window function & Hann & -- \\
Mel filterbank size & $128$ & $f_{\min}=0$, $f_{\max}=8000\,\mathrm{Hz}$ \\
Time frames $T_f$ & $1000$ & $100\,\mathrm{frames/s}\times 10\,\mathrm{s}$ \\
Input channels & $7$ & mel channels $(W,Y,Z,X)$ + IV channels $(IV_x,IV_y,IV_z)$ \\
IV formula & $\mathrm{IV}_d=\mathrm{Re}(W\overline{X_d})/(|W|^2+\varepsilon)$ & $\varepsilon=10^{-8}$, clamp to $\pm 10$ after mel extraction \\
W-channel mean & $15.41663$ & BEATs pretrain statistic \\
W-channel std & $6.55582$ & BEATs pretrain statistic \\
SpecAugment (W only) & $2\times$ time mask, $2\times$ freq mask & Training only \\
\bottomrule
\end{tabularx}
\caption{FOA input and feature extraction parameters.}
\label{tab:v13d-foa-input}
\end{table*}

\subsection{SO-Encoder Training}
We use a two-stage training strategy for the SO-Encoder.
In the first stage, we perform isolated class learning.
We freeze spatial losses and only optimize the backbone trunk, adapter, and class head, so that the event semantic branch becomes stable before spatial gradients are introduced.
In the second stage, we linearly warm up spatial losses and jointly optimize class, active, DOA, and distance supervision.
We use cosine decay learning rate scheduling to reduce early interference from spatial fusion.
We also balance the ratio of real and simulated data during training by oversampling the real data, with a real-to-simulated ratio of $1:2$. 
We perform random cropping of at most $10\,\mathrm{s}$ during training, and use the full clip during validation and testing.
The details of the training configuration are summarized in Table~\ref{tab:v13d-training}.

\begin{table*}[h]
\centering
\small
\begin{tabular}{lll}
\toprule
\textbf{Group} & \textbf{Parameter} & \textbf{Value} \\
\midrule
\multirow{5}{*}{Loss weights}
 & $\lambda_{\text{frame\_class}}$       & $1.0$ ($63$-way cross-entropy) \\
 & $\lambda_{\text{frame\_activity}}$    & $1.0$ (Top-$K$ rank, replaces BCE) \\
 & $\lambda_{\text{frame\_direction}}$   & $1.0$ ($1 - \cos(\hat{\mathbf{d}}, \mathbf{d})$) \\
 & $\lambda_{\text{frame\_distance}}$    & $1.0$ (smooth-$\ell_1$) \\
\midrule
\multirow{4}{*}{Top-$K$ rank loss}
 & $K$ & $4$ \\
 & \texttt{frame\_activity\_loss\_type}  & \texttt{topk\_rank} \\
 & Margin $m$                            & $2.0$ \\
 & BCE anchor weight                     & $0.1$ \\
\midrule
\multirow{4}{*}{Two-phase schedule}
 & LR warmup, epochs $0$--$2$            & Spatial weight $0$, LR $0 \to 1.5{\times}10^{-5}$ \\
 & Class-only warmup, epochs $3$--$7$    & Spatial weight $0$, cosine decay \\
 & Spatial ramp, epochs $8$--$9$         & Spatial weight $0 \to 1$, cosine decay \\
 & Full joint training, epochs $10$--$24$ & Spatial weight $1$, cosine decay to $7.5{\times}10^{-7}$ \\
\midrule
\multirow{5}{*}{Optimizer}
 & Optimizer                             & \texttt{AdamW} \\
 & $(\beta_1, \beta_2)$                  & $(0.9, 0.999)$ \\
 & $\varepsilon$                         & $10^{-8}$ \\
 & Weight decay                          & $0.01$ \\
 & Gradient clipping (global $\ell_2$)   & $1.0$ \\
\midrule
\multirow{5}{*}{LR schedule}
 & LR schedule                           & Linear warmup and cosine decay \\
 & Peak LR                               & $1.5 \times 10^{-5}$ \\
 & LR warmup epochs                      & $3$ \\
 & Cosine decay epochs                   & $22$ \\
 & Min LR ratio                          & $0.05$ \\
\midrule
\multirow{6}{*}{Training scale}
 & Total epochs                          & $25$ \\
 & GPUs                                  & $8\times$ A100 \\
 & Batch size (per GPU / total)          & $8 / 64$ \\
 & Numerical precision                   & \texttt{FP32} \\
 & Data loader workers / GPU             & $8$ \\
 & Distributed framework                 & \texttt{torchrun} + \texttt{DDP} \\
\midrule
\multirow{4}{*}{EMA}
 & \texttt{use\_ema}                     & \texttt{true} \\
 & EMA decay                             & $0.9995$ \\
 & EMA start epoch                       & $3$ \\
\bottomrule
\end{tabular}
\caption{SO-Encoder training configuration.}
\label{tab:v13d-training}
\end{table*}

\noindent The Top-$K$ rank activity loss is
\[
\begin{aligned}
\mathcal{L}_{\text{rank}}
&= \frac{1}{|P|} \sum_{(i,j) \in P} \max\!\left(0,\; m + \ell_j - \ell_i\right), \\
\mathcal{L}_{\text{act}}
&= \mathcal{L}_{\text{rank}} + 0.1 \cdot \mathcal{L}_{\text{BCE}}.
\end{aligned}
\]
where $P$ is the set of (active slot $i$, inactive slot $j$) pairs within each frame.

\subsection{Spatial-Omni Model Training}
\label{app: spatial_llm_training_details}
The training of Spatial-Omni models follows a three-stage strategy, allowing the spatial tokens to gradually adapt to the LLM without disrupting the original audio and language capabilities.
We train the projector in the first stage, the projector and LLM LoRA parameters in the second stage, and the SO-Encoder, projector, and LLM LoRA parameters in the third stage.
The specific parameter counts and learning rate settings are shown in Table~\ref{tab:stage_schedule}. 
Maximum audio length is set to $20\,\mathrm{s}$ during training.
Each stage continues training from the best checkpoint of the previous stage.
Table~\ref{tab:shared_hparams} summarizes the hyperparameters shared across all three stages of training.

\begin{table*}[h]
  \centering
  \small
  \resizebox{\textwidth}{!}{
  \begin{tabular}{c l l c c c c}
  \toprule
  \textbf{Stage} & \textbf{Trainable Modules} & \textbf{Base LR} & \textbf{Projector LR} & \textbf{LoRA LR} &
  \textbf{Encoder LR} & \textbf{Epochs} \\
  \midrule
  1 & Projector ($\sim 17\,\mathrm{M}$)
    & $5{\times}10^{-5}$ & $1{\times}10^{-4}$ & --                 & --                 & $2$ \\
  2 &  Projector + LoRA on \texttt{q\_proj,k\_proj,v\_proj,o\_proj} ($\sim 60\,\mathrm{M}$)
    & $5{\times}10^{-5}$ & $3{\times}10^{-5}$ & $5{\times}10^{-5}$ & --                 & $3$ \\
  3 &  Projector + LoRA + SO-Encoder ($\sim 162\,\mathrm{M}$)
    & $3{\times}10^{-5}$ & $1{\times}10^{-6}$ & $3{\times}10^{-5}$ & $1{\times}10^{-6}$ & $3$ \\
  \bottomrule
  \end{tabular}
  }
  \caption{Spatial-Omni three-stage training schedule.}
  \label{tab:stage_schedule}
\end{table*}

\begin{table}[h]
  \centering
  \small
  \begin{tabularx}{\columnwidth}{@{}lX@{}}
  \toprule
  \textbf{Hyperparameter} & \textbf{Value} \\
  \midrule
  \multicolumn{2}{@{}l}{Batching} \\
  Per-GPU batch size      & $2$ \\
  Gradient accumulation steps                & $3$ \\
  GPUs (single node)      & $8$ \\
  Effective global batch size                & $48$ \\
  \midrule
  \multicolumn{2}{@{}l}{Optimization} \\
  Optimizer               & \texttt{AdamW} \\
  LR schedule             & Linear warmup, cosine decay \\
  Warmup ratio       & $0.03$ \\
  Weight decay      & $0.01$ \\
  Gradient clipping & $1.0$ \\
  \midrule
  \multicolumn{2}{@{}l}{Precision} \\
  Compute dtype            & {bf16} \\
  Spatial modules dtype    & {fp32} \\
  \midrule
  \multicolumn{2}{@{}l}{LoRA (Stages 2 and 3)} \\
  Rank & $16$ \\
  $\alpha$          & $32$ \\
  Dropout            & $0.05$ \\
  Target modules          & \texttt{q,k,v,o\_proj} \\
  \midrule
  \multicolumn{2}{@{}l}{Data} \\
  Audio & $4$-ch FOA, $16\,\mathrm{kHz}$, $\le 20\,\mathrm{s}$ \\
  Spatial-token rate      & $2.5\,\mathrm{Hz}$ \\
  \bottomrule
  \end{tabularx}
  \caption{Hyperparameters shared by all three stages of Spatial-Omni model training.}
  \label{tab:shared_hparams}
\end{table}

\section{Benchmark Details}
\label{appendix:benchmark}
SO-Bench uses an independent spatial QA test set to avoid overlap with training data.
The benchmark covers three groups of tasks.
Basic detection and localization include Detect Source, Detect Time, Estimation of Azimuth, Estimation of Elevation, Estimation of Distance.
Spatial understanding includes Identify Source by DoA, and Identify Source by Location, Relative Left Right, Comparison Elevation, Comparison Distance, and Onset from Location.
Complex spatial reasoning includes Classify Motion, Count Sources, Multi Hop, Spatial Temporal caption, and Speech Content Recognition.

For evaluation, we use task-specific metrics.
Detect Source is evaluated with event F1:
\[
\mathrm{F1} = \frac{2 \cdot \mathrm{Precision} \cdot \mathrm{Recall}}{\mathrm{Precision} + \mathrm{Recall}}.
\]
Detect Time is evaluated with time-span IoU:
\[
\mathrm{IoU}_{\mathrm{time}} =
\frac{\left|[t_s^{p}, t_e^{p}] \cap [t_s^{g}, t_e^{g}]\right|}
{\left|[t_s^{p}, t_e^{p}] \cup [t_s^{g}, t_e^{g}]\right|},
\]
where $[t_s^{p}, t_e^{p}]$ and $[t_s^{g}, t_e^{g}]$ denote the predicted and ground-truth time spans.
For tolerance-based estimation and classification tasks, we report accuracy:
\[
\mathrm{Acc} = \frac{N_{\mathrm{correct}}}{N_{\mathrm{total}}}.
\]
For azimuth, elevation, and distance estimation, $N_{\mathrm{correct}}$ counts predictions within tolerances of $20^\circ$, $10^\circ$, and 0.5 m, respectively.
For Onset from Location, a prediction is correct when the onset error is within 0.4 seconds.
Identify Source by DoA, Identify Source by Location, Relative Left-Right, Comparison between Elevation, Comparison between Distance, Classify Motion, Count Sources, and Multi Hop use exact-match or choice accuracy according to their answer format.
For Spatial Temporal caption, we compute F1 over the required spatial attributes and event classes.
For Speech Content Recognition, we compute WER between the predicted transcript and the ground-truth transcript for the specified direction:
\[
\mathrm{WER} = \frac{S + D + I}{N},
\]
where $S$, $D$, and $I$ are the numbers of substitutions, deletions, and insertions, and $N$ is the number of words in the reference transcript.

\section{Details of Baselines}
\label{app: baseline}

\subsection{Encoder Baselines}
\label{app: encoder_baselines}
Publicly available spatial audio understanding models include SELDNet ~\cite{adavanne2018sound}, the DCASE baseline series ~\cite{Adavanne2019_DCASE}, and models like Spatial-AST ~\cite{zheng2024bat}.
SELDNet is based on CNN and GRU modules to extract simple spatial audio features, supporting multi-source event detection and localization, but its performance is limited in complex scenarios.
The DCASE challenge series provides baseline models for spatial audio understanding, consisting of CNN and Transformer modules, and introduces the Multi-ACCDOA output format, supporting multi-source event detection and localization. However, its performance is also limited in complex scenarios, especially in tasks with increased sound event categories and class imbalance, where it performs poorly even with the addition of a Sound Event Detect head.
In the DCASE 2024 challenge, except for the best team which achieved an F20 score of 54.4\%, the rest of the teams did not exceed an F20 score of 30\%. Moreover, it was limited to the STARSS23 dataset. We reproduced the best team's model, but due to the lack of pretrained checkpoints and augmented datasets, we did not achieve the performance level published in their paper.
Spatial-AST is designed based on AudioMAE, using a Vision Transformer architecture and concatenating class tokens, making it a representative model in the field of binaural audio understanding. The spatial tokens extracted by Spatial-AST can provide a certain level of spatial understanding when integrated into the later LLaMA model.

\subsection{Spatial Audio LLM Baselines}
\label{app: spatial_llm_baselines}
We design multiple baselines to compare and validate the effectiveness of our framework.
BAT is currently the only open-source binaural audio LLM model, supporting detection and localization of one to two sound sources. We integrate Spatial-AST into the LLaMA framework to achieve spatial audio understanding.
The IV baseline directly extracts three sets of IV features between the W channel and X/Y/Z channels, and performs simple dimensionality reduction through pooling in the frequency and time domains before inputting them into the LLM as spatial tokens. This is similar to the way IV features are concatenated into the audio encoder in ~\cite{tang2024can}.
The Neural IV baseline is based on the learnable IV features implemented in ~\cite{liu2026jaeger}. The 3-channel IV features are processed through two layers of Conv2D, downsampled in the time dimension using pooling, and then projected to the LLM's spatial token dimension through LayerNorm and MLP layers.  
Given a 4-channel FOA waveform sampled at $16\,\mathrm{kHz}$, we run the preprocessing and feed the three IV channels into a two-layer 2-D CNN ($3\to32\to64$, $3{\times}3$ kernels, GELU, $T$ and $M$ unchanged),
average over the mel axis, downsample the 50 Hz time axis to 2.5 Hz with adaptive average pooling, and pass the result through LayerNorm and a two-layer MLP ($64\to256\to256$) scaled by $0.02$ and clipped to $[-1,1]$. A second MLP projector ($256\to512\to3584$) lifts each spatial token to the Qwen2.5-Omni-7B hidden size and the resulting 50 tokens are masked-scattered into the LLM's input sequence at the <|spatial|> placeholder positions. 
The CNN, token head, and MLP layers together contain $\sim$5 M trainable parameters.
The Zero-Spatial baseline simulates the case of monaural audio without spatial information by feeding a null spatial token into the LLM, to verify the contribution of spatial tokens to spatial relationship understanding and reasoning.
Besides the baselines mentioned in the main text, we also attempted to integrate the DCASE baseline model trained on real data into the Qwen2.5-Omni model. The trained model achieved a level close to Neural IV in basic sound event detection and localization tasks, but performed weaker in more complex spatial relationship understanding and reasoning tasks, showing no significant improvement. This demonstrates the limited spatial performance capability of the DCASE baseline encoder.

\section{Supplementary Ablations and Results}
\label{app: supplementary_results}

\begin{table*}[ht]
    \centering
    \resizebox{\textwidth}{!}{
    \begin{tabular}{l|ccccc|cccccc|ccccc}
        \toprule
        \multirow{2}{*}{\textbf{Model}} & \multicolumn{5}{c|}{\textbf{Basic Detection and Estimation}} & \multicolumn{6}{c|}{\textbf{Spatial Relation Understanding}} & \multicolumn{5}{c}{\textbf{Complex Reasoning and Semantics}} \\
        & \textbf{DS} & \textbf{DT} & \textbf{EAzi} & \textbf{EEle} & \textbf{EDis} & \textbf{IS-DoA} & \textbf{IS-Loc} & \textbf{RLR} & \textbf{CEle} & \textbf{CDis} & \textbf{OL} & \textbf{CM} & \textbf{CS} & \textbf{MH} & \textbf{ST} & \textbf{SC$\downarrow$} \\
        \midrule
        Easy stage1   & 10.50 & 59.51 & 19.67 & 34.04 & 50.48 & - & -     & -     & -     & -     & -     & -     & -     & -     & -     & -    \\
        Easy stage2   & 36.89 & 73.74 & 35.14 & 47.23 & 58.59 & - & -     & -     & -     & -     & -     & -     & -     & -     & -     & -    \\
        Easy stage3   & 38.19 & 74.29 & 44.14 & 59.37 & 60.84 & - & -     & -     & -     & -     & -     & -     & -     & -     & -     & -    \\
        Easy + Medium & 47.14 & 77.02 & 55.10 & 66.21 & 76.19 & 59.41  & 53.57 & 55.85 & 61.26 & 60.36 & 77.97 & -     & -     & -     & -     & -    \\
        \midrule
        Full & {51.72} & 87.01 & 76.38 & 76.92 & 77.77 &{68.06} &  {66.81} & 72.97 & {81.68} & 57.95 &  {86.65} &  {67.74} & 47.05 &  {39.93} &  {57.02} & 75.94 \\
        \bottomrule
    \end{tabular}
    }
    \vspace{0.1cm}
    \caption{Per-stage results for Spatial-Omni training. Easy for Basic Detection and Estimation, Medium for Spatial Relation Understanding, and Hard for Complex Reasoning and Semantics.}
    \label{tab:stage_results}
\end{table*}

\subsection{Per-Task Analysis}
\label{app: per_task_analysis}
For basic sound event detection and time detection tasks, the base model has some capability, and our SO-Encoder outputs frame-level spatial representations, which can enhance temporal localization ability.
For angle estimation tasks, the median error for random guessing is about 90 degrees for azimuth and about 45 degrees for elevation; if we calculate accuracy using an angle tolerance, the random baseline is about 11.11\%. Our model has significant improvements in both tasks, while LALMs that only accept monaural input and the base Omni LLM perform close to random levels.
For distance detection tasks, our model has significant improvements, and there is also a significant improvement in absolute distance error.
In the two localization event detection tasks, our model achieve strong performance. BAT's QA design is specifically trained for this type of problem, without training for more complex spatial relationship understanding and reasoning problems, while our model has a more balanced performance in these types of problems.
In comparison tasks, our model has improvements, indicating that in multi-source scenarios, our model has stronger discriminative ability. This limited improvement also indicates that the current event detection and track assignment capabilities of the SO-Encoder still limit downstream multi-source spatial relationship reasoning.
The good performance of the neural IV baseline on comparison problems is consistent with the conclusions in JAEGER ~\cite{liu2026jaeger}, where in comparison scenarios with only two sound sources, learnable IV features have good effects on spatial relationship understanding.
The improvement in the onset from location task indicates that our model can better utilize spatial information for temporal localization, effectively outputting frame-level spatial information.
In the more difficult motion judgment relationship problem, our model can better utilize frame-level spatial information to determine the motion state of sound events.
The count source task requires a high level of spatial audio understanding. In the work of ~\cite{you2026world}, counting the number of sources is also difficult. Our method still does not achieve good performance, while BAT's model only supports detection of at most 2 sound sources, so it can perform well on counting source tasks with fewer than 2 sources, but cannot handle higher numbers of sources.
The multi-hop task requires the model to understand the spatial relationships between multiple sound events and perform complex reasoning. Our method has significant improvements in this type of problem, indicating that our model can better utilize spatial information for complex spatial relationship understanding and reasoning.
The improvement in Spatial Temporal captioning indicates that our model can better utilize spatial information for spatial event description and understanding, extracting the relationship between space and semantics compared to the baseline.
The Speech content task is evaluated using WER, where a lower value indicates more accurate recognition of speech content in the specified direction. In this task, our model can recognize the specified directional speech content in overlapping speech, with improvement compared to the base model.

\begin{table}[htb]
    \centering
    \resizebox{\columnwidth}{!}{
    \begin{tabular}{l|ccc}
        \toprule
        \textbf{Stage} & \textbf{Median Azi Err} & \textbf{Median Ele Err} & \textbf{Median Dis Err} \\
        \midrule
        Easy stage1   & 55.6 & 15.1 & 0.54 \\
        Easy stage2   & 34.8 & 10.3 & 0.48 \\
        Easy stage3   & 26.6 & 8.3 & 0.45 \\
        Easy + Medium & 16.9 & 6.7  & 0.41 \\
        Full          & {7.6} & {4.0}  & {0.40} \\
        \bottomrule
    \end{tabular}
    }
    \vspace{0.1cm}
    \caption{Per-stage results for basic estimation tasks in Spatial-Omni training. Angular errors are reported in degrees, and distance errors in meters.}
    \label{tab:stage_results_degree}
\end{table}

\begin{table*}[htb]
    \centering
    \resizebox{\linewidth}{!}{
    \begin{tabular}{l|ccc}
        \toprule
        \textbf{Model} & \textbf{Total Parameters} & \textbf{Peak Inference GPU Memory ($\mathrm{GB}$)} & \textbf{End-to-End Inference Speed ($\mathrm{s}$)} \\
        \midrule
        Qwen2.5-Omni & $8.93\,\mathrm{B}$ & $16.99$ & $1.85$ \\
        SO-7B-neuiv & $8.93\,\mathrm{B}$ & $16.99$ & $1.88$ \\
        SO-7B & $9.09\,\mathrm{B}$ & $17.63$ & $1.94$ \\
        \bottomrule
    \end{tabular}
    }
    \vspace{0.1cm}
    \caption{Total parameters, peak inference GPU memory, and end-to-end inference speed of Qwen2.5-Omni, SO-7B-neuiv, and SO-7B.}
    \label{tab:parameters}
\end{table*}
\begin{table*}[htb]
    \centering
    \resizebox{\textwidth}{!}{
    \begin{tabular}{l|cc|cc|cc|cc|cc|cc}
        \toprule
        \multirow{2}{*}{\textbf{Stage}} & \multicolumn{2}{c|}{\textbf{Det F1 (\%)}} & \multicolumn{2}{c|}{\textbf{Time IoU (\%)}} & \multicolumn{2}{c|}{\textbf{Az Acc (\%)}} & \multicolumn{2}{c|}{\textbf{Az Err ($^\circ$)}} & \multicolumn{2}{c|}{\textbf{El Acc (\%)}} & \multicolumn{2}{c}{\textbf{El Err ($^\circ$)}} \\
        & \textbf{B1} & \textbf{B4} & \textbf{B1} & \textbf{B4} & \textbf{B1} & \textbf{B4} & \textbf{B1} & \textbf{B4} & \textbf{B1} & \textbf{B4} & \textbf{B1} & \textbf{B4} \\
        \midrule
        Easy Stage 1 & 10.50 & 10.61 & 59.51 & 62.23 & 19.67 & 20.87 & 55.6 & 62.0 & 34.04 & 32.98 & 15.1 & 15.9 \\
        Easy Stage 2 & 36.89 & 35.70 & 73.74 & 73.54 & 35.14 & 33.78 & 34.8 & 37.5 & 47.23 & 49.93 & 10.3 & 10.0 \\
        Easy Stage 3 & 38.19 & 37.52 & 74.29 & 73.76 & 44.14 & 41.59 & 26.6 & 30.0 & 59.37 & 62.52 & 8.3 & 7.5 \\
        \bottomrule
    \end{tabular}
    }
    \vspace{0.1cm}
    \caption{Beam ablation study comparing Beam=1 (greedy) and Beam=4 decoding across Spatial-Omni training stages. B1 denotes Beam=1, B4 denotes Beam=4, and Err denotes median degree error.}
    \label{tab:beam_ablation}
\end{table*}

\subsection{Per-Stage Results for Spatial-LLM Training}
We employ a staged training strategy. We first train the model on easy QA data to learn basic spatial understanding capabilities, and then gradually add medium and hard stage data to further enhance the model's abilities. In the easy stage, we first train the projector, and then unfreeze the LoRA parameters of the LLM for training with a stable projector initialization. In the final stage, we jointly train the full SO-Encoder, projector, and LLM.
The results for each stage are shown in Table~\ref{tab:stage_results}. We can see that training in the easy stage already enables the model to learn basic spatial understanding capabilities, while training in the medium and hard stages further enhances the model's abilities, especially in complex scenarios.
We further measure median degree error for azimuth and elevation estimation, and absolute distance error for distance estimation for each stage, shown in Table~\ref{tab:stage_results_degree}. 
Although harder-stage questions do not directly supervise angle or distance estimation, the staged training still improves the learned spatial representations.

\subsection{Ablation on Mix Data Training}
\label{app:mix_data}
As the training data for the Qwen series has not been open-sourced, we supplemented it with a portion of the open-source single-channel QA data from Audio-Flamingo3. 
The data sources include the QA subsets of CochlScene, Audio\_SL, MusicCaps, FSD50k, and UrbanSound8K, totaling $130$k QA pairs. 
During training, we randomly sampled 20\% of the spatial training sets and mixed them with the aforementioned single-channel dataset, resulting in an overall data ratio of $1:3$ for $1$ epoch. 
The single-channel audio is input into SO-Encoder in the form of $[W, 0, 0, 0]$ for feature extraction, allowing the SO-Encoder to learn the distribution of non-spatial features and output a null spatial token feature representation to distinguish between single-channel and spatial audio inputs. 
The resulting model, SO-7B(MIX), shows improvements in Sound, Music, and Speech capabilities on MMAU and MMAU-Pro, while still enhancing spatial audio capabilities on MMAU-Pro. 
Although it did not reach the capabilities of the original base Omni model due to limited data, it demonstrates the feasibility of mixed training and the ability of the spatial model to learn from single-channel data.
Further gains may be obtained if the original Qwen training data is available for joint training.  

The mixed data training did not unfreeze the original audio encoder, but by using single-channel input, it allows the model to learn when to utilize single-channel audio tokens and when to utilize spatial tokens, as well as learning the feature representation of a null spatial token when there is no spatial audio. 
This is helpful for improving the model's capabilities, as it learns to better combine single-channel and spatial features for understanding and reasoning.
The model also learns to adapt to the distribution of the added spatial modality, resulting in improvements in metrics for single-channel tasks.

\subsection{General-Audio Retention Results}
\label{app:mmaus}

\begin{table*}[t]
    \centering
    \small
    \setlength{\tabcolsep}{2.3mm}
    \resizebox{\textwidth}{!}{
    \begin{tabular}{l|cccc|ccccc}
        \toprule
        \multirow{2}{*}{\textbf{Model}} & \multicolumn{4}{c|}{\textbf{MMAU}} & \multicolumn{5}{c}{\textbf{MMAU-Pro}} \\
        & \textbf{avg} & \textbf{sound} & \textbf{music} & \textbf{speech} & \textbf{avg} & \textbf{sound} & \textbf{music} & \textbf{speech} & \textbf{spatial} \\
        \midrule 
        Random & 26.00 & 26.72 & 24.55 & 26.72 & 23.40 & 28.30 & 26.10 & 29.40 & 21.20 \\
        \midrule
        Phi-4-mm & 62.50 & 64.56 & 68.56 & 54.35 & 44.13 & 36.39 & 49.58 & 46.91 & 32.62 \\
        SO-4B & 9.00 & 14.41 & 5.09 & 7.51 & 8.55 & 7.55 & 7.40 & 11.00 & 11.08 \\
        \rowcolor{green!10}
        SO-4B (MIX) & 58.60 & 62.76 & 62.57 & 50.45 & 42.23 & 37.82 & 45.98 & 44.56 & 32.31 \\
        \midrule
        Audio Flamingo 3 & 67.70 & 70.87 & 75.45 & 56.76 & 57.31 & 57.31 & 62.02 & 52.64 & 54.77 \\
        {SO-AF3} & 33.20 & 47.15 & 27.84 & 24.62 & 21.43 & 16.71 & 20.38 & 19.19 & 43.08 \\
        \rowcolor{green!10}
        {SO-AF3(MIX)} & 68.20 & 71.17 & 76.65 & 56.76 & 58.23 & 58.07 & 62.27 & 52.64 & 61.23 \\
        \midrule
        Kimi-Audio & 65.40 & 72.97 & 64.97 & 58.26 & 56.43 & 58.26 & 57.40 & 53.31 & 57.85 \\
        {SO-Kimi} & 45.60 & 54.05 & 41.02 & 41.74 & 35.60 & 35.63 & 33.92 & 31.20 & 48.31 \\
        \rowcolor{green!10}
        {SO-Kimi(MIX)} & 54.20 & 66.67 & 52.69 & 43.24 & 43.69 & 49.28 & 44.43 & 37.37 & 38.46 \\
        \midrule
        Qwen2.5-Omni & 69.99 & 73.27 & 70.65 & 66.06 & 60.36 & 61.98 & 64.66 & 62.51 & 31.07 \\
        SO-7B & 60.20 & 68.17 & 55.69 & 56.76 & 41.83 & 38.26 & 50.42 & 41.64 & 48.16 \\
        \rowcolor{green!10}
        SO-7B (MIX) & 66.20 & 69.37 & 64.37 & 64.86 & 51.15 & 52.19 & 56.91 & 52.64 & 47.24 \\
        \midrule
        Qwen3-Omni (ours) & 68.40 & 69.97 & 71.56 & 63.66 & 62.84 & 47.09 & 72.99 & 72.62 & 36.62 \\
        SO-30B & 65.00 & 64.56 & 69.16 & 61.26 & 52.13 & 46.99 & 60.16 & 49.49 & 44.92 \\
        \rowcolor{green!10}
        SO-30B (MIX) & 69.80 & 70.27 & 75.15 & 63.96 & 56.23 & 52.63 & 62.02 & 56.90 & 46.15 \\
        \bottomrule
    \end{tabular}}
    \caption{Raw MMAU and MMAU-Pro retention results underlying Figure~\ref{fig:mmau_retention}. The spatial column denotes MMAU-Pro's spatial-audio category.}
    \label{tab:mmaus_appendix}
\end{table*}

\subsection{Inference Speed, Token Rate, and Memory Usage}
\label{app: inference_speed}
We compared the original Qwen2.5-Omni and SO-7B with the SO-Encoder in terms of inference speed, GPU memory usage, and parameter count.
Qwen2.5-Omni has $8.93\,\mathrm{B}$ parameters, SO-7B-neuiv adds only about $5\,\mathrm{M}$ parameters through the CNN modules, resulting in negligible parameter overhead.
SO-7B has $9.09\,\mathrm{B}$ parameters due to the addition of the SO-Encoder and projector.
Qwen2.5-Omni has a peak GPU memory usage of $16.99\,\mathrm{GB}$ during inference, and SO-7B has an increased peak GPU memory usage due to the addition of the SO-Encoder and projector, but it is still within an acceptable range.
For inference speed, we tested the inference time with $64$ new tokens. The addition of new spatial tokens does not significantly increase inference time.
The results are shown in Table~\ref{tab:parameters}.

We train SO-7B on $8\times$ A100 NVIDIA GPUs for $576$ GPU hours and SO-30B on $8\times$ H20 GPUs for $768$ GPU hours.

\subsection{Ablation on Decoding}
When answering questions related to angles, since the LLM outputs token by token, there may be an issue that the model simply memorizes templated angle responses. To verify this, we design an ablation experiment focusing on detect and angle questions for the 3 easy QA stages with beam ablation.
The ablation experiment is set up with beam=4, while the rest of the settings are the same as the default experiment. The results are shown in Table~\ref{tab:beam_ablation}. In the greedy setting, we obtained 42.9\% integer angle answers, while in the beam=4 setting, we only had 10.5\%, which is consistent with the real distribution. Beam=4 does not substantially change detection or time-localization performance, while azimuth errors become slightly larger and elevation errors remain comparable or slightly better in later stages.
Our model does not fall into templated angle answering. The elevation results are better because we simulate a realistic data distribution where most elevation values are within $\pm 30^\circ$, so the model's predictions are biased toward around $0^\circ$ and still achieve good accuracy within a 10-degree tolerance.
In contrast, azimuth is uniformly distributed within a 360-degree range, which can better reflect the learning ability. Our model indeed learns spatial information from the spatial token.

Current open-source model evaluations typically use greedy decoding. To align with these implementations, we report greedy decoding in the main results and provide beam-search results as a decoding ablation.

\section{LLM Prompts}
We use LLMs for QA generation and paraphrasing. We present the prompts used for LLM QA generation and paraphrasing below.

\begin{promptbox}[label={lst:qa-generation-prompt}]{Prompt used for QA generation}
You are given the metadata of a spatial audio clip, including sound events, active time intervals, azimuth, elevation, distance, and motion information.

Your task is to generate diverse spatial audio question-answering pairs.

Requirements:
1. The question must be answerable from the provided metadata.
2. The question should focus on spatial audio understanding, such as source detection, direction estimation, distance estimation, motion analysis, spatial relation reasoning, or speech content under spatial conditions.
3. Do not mention metadata fields explicitly in the question.
4. The answer should be concise and factual.
5. If multiple sound sources are present, ensure that the question clearly specifies the target source, time interval, or spatial condition.

Metadata:
{metadata}

Please generate {num_questions} question-answer pairs in JSON format.
\end{promptbox}

\begin{promptbox}[label={lst:qa-paraphrase-prompt}]{Prompt used for QA paraphrasing}
You are given a spatial audio question-answer pair.

Your task is to rewrite the question into a natural and diverse form while
preserving its original meaning.

Requirements:
1. Do not change the answer.
2. Do not introduce new spatial information.
3. Keep the rewritten question answerable from the same audio clip.
4. Avoid overly formal or repetitive wording.
5. Preserve the target source, time interval, and spatial relation if they are mentioned in the original question.

Original question:
{question}

Answer:
{answer}

Return only the rewritten question.
\end{promptbox}

\section{Licenses and Availability}
We respect the original licenses of all referenced artifacts and do not redistribute them. 
This work uses publicly available datasets. 
We do not redistribute any third-party audio content. 
Users must obtain the original datasets from their respective providers and comply with the original licenses/terms of use. 
We will release data under CC BY 4.0, subject to the original dataset terms. 
Our codebase may depend on third-party libraries; these components remain under their respective licenses. 
Any external assets (e.g., pretrained backbones or evaluation tools) are used in accordance with their original licensing terms.

\section{Use of AI Assistants}
We used AI-based writing assistant during manuscript preparation solely for language polishing, including grammar checking, spelling correction, and improving clarity and readability of the text. 
All technical claims, experimental procedures, and interpretations were produced and verified by the authors.

\end{document}